\documentclass[11pt,twoside,letter]{article}
\usepackage{graphicx}
\usepackage{fancyhdr}
\usepackage{amsmath, amsthm, amssymb}
\usepackage[T1]{fontenc}
\usepackage[boldsans]{concmath}
\usepackage{lscape}
\usepackage{fullpage}
\usepackage{csquotes}
\usepackage{cite}
\usepackage{epstopdf}
\usepackage{epsfig}
\usepackage{algorithm, algorithmic}
\usepackage[lofdepth,lotdepth]{subfig}

\theoremstyle{plain}

\newtheorem{thm}{Theorem}

\theoremstyle{definition}
\newtheorem{dfn}[thm]{Definition}

\theoremstyle{remark}
\newtheorem{rem}[thm]{Remark}
\newtheorem{per}[thm]{Proposition}

\usepackage[headsep=18pt]{geometry} 
\begin{document}

%
%
\title{\bf\vspace{-18pt} Non-Convex Compressed Sensing Using Partial Support Information}

%
%


\author{Navid Ghadermarzy \\ \small Department of Mathematics, University of British Columbia \\
\small Vancouver, BC, Canada \\ \small navidgh@math.ubc.ca \\
\\
Hassan Mansour \footnote{This work was conducted while Hassan Mansour was a postdoctoral research fellow in the Mathematics Department at the University of British Columbia.} \\ \small Mitsubishi Electric Research Laboratories \\
\small Cambridge, MA 02139, USA \\ \small mansour@merl.com \\
\\
{\"O}zg{\"u}r Y{\i}lmaz \\ \small Department of Mathematics, University of British Columbia\\
\small Vancouver, BC, Canada \\ \small oyilmaz@math.ubc.ca \\
}

%
%
\date{}

%
%
\maketitle
\thispagestyle{fancy}

%
%
\markboth{\footnotesize \rm \hfill N. Ghadermarzy and H. Mansour and O. Y{\i}lmaz \hfill \vspace{39pt}}
{\footnotesize \rm \hfill Non-Convex Compressed Sensing Using Partial Support Information \hfill \vspace{39pt}}

%
%

\begin{abstract}
In this paper we address the recovery conditions of weighted $\ell_p$ minimization for signal reconstruction from compressed sensing measurements when partial support information is available. We show that weighted $\ell_p$ minimization with $0<p<1$ is stable and robust under weaker sufficient conditions compared to weighted $\ell_1$ minimization. Moreover, the sufficient recovery conditions of weighted $\ell_p$ are weaker than those of regular $\ell_p$ minimization if at least $50\%$ of the support estimate is accurate. We also review some algorithms which exist to solve the non-convex $\ell_p$ problem and illustrate our results with numerical experiments.
\vspace{5mm} 
\vspace{3mm}
\end{abstract}

%
%

\section{Introduction}
Compressed sensing is a data acquisition technique for efficiently recovering sparse signals from seemingly incomplete and noisy linear measurements. There are many applications where the target signals admit sparse or nearly sparse representations in some transform domain. For example, natural images are nearly sparse in discrete cosine transform domain (DCT) and in the wavelet domain. Similarly audio signals are approximately sparse in short time Fourier domain.\newline
Compressed sensing is especially promising in applications where taking measurements is costly, e.g., hyperspectral imaging \cite{hyperspectral_fowler}, as well as in applications where the ambient dimension of the signal is very large, i.e., medical \cite{lustig2007sma} and seismic imaging \cite{simply_denoise}.\newline
Define $\Sigma_k^N:= \{u \in \mathbb{R}^N : \| u \|_0 \leq k\}$ to be the set of all $k$-sparse vectors in $\mathbb{R}^N$---$\| u \|_0$ denotes the number of non-zero components of $u$. Let $x \in \Sigma_k^N$ and assume that $y \in \mathbb{R}^n$, the vector of $n$ linear and potentially noisy measurements of $x$, is acquired via $y:= Ax + e$ where $e$ denotes the noise in our measurements with $\|e\|_2 \leq \epsilon$. Here A is an $n\times N$ measurement matrix with $n \ll N$. We wish to recover $x$ from $y$ by solving a sparse recovery problem. This entails finding the sparsest vector $\hat{x}$ that is feasible, i.e., $\| A\hat{x}-y \| \leq \varepsilon$. In the noise free case, i.e., $\epsilon=0$, the decoder $\bigtriangleup_0: \mathbb{R}^{n\times N} \times \mathbb{R}^n \mapsto \mathbb{R}^N$ is defined as
\begin{equation} \label{l0}
    \begin{aligned}
	\bigtriangleup_0(A,y) := \underset{z \in \mathbb{R}^N} {\text{argmin}} \ \lvert\lvert z \lvert\lvert_0   \ \ 
	                                   \text{s.t.} \ \  Az=y. 
	                                   \end{aligned}
\end{equation}
It was proved, e.g., in \cite{Donoho03}, that if $n>2k$ and $A$ is in general position, i.e., any collection of $n$ columns of $A$ is linearly independent, then $\bigtriangleup_0(A,y)=x$. However, (\ref{l0}) is a combinatorial problem which becomes intractable as the dimensions of the problem increase. Therefore, one seeks to modify the optimization problem so that it can be solved with methods that are more tractable than combinatorial search.\newline
Donoho \cite{Donoho2006_CS} and Cand{\'e}s, Romberg, and Tao \cite{CRT05} showed that if $A$ obeys a certain \enquote{restricted isometry property}, solving a convex relaxation to the $\ell_0$ problem can stably and robustly recover x from 
measurements $y= Ax + e$. More precisely, $\bigtriangleup_1: \mathbb{R}^{n\times N} \times \mathbb{R}^n \times \mathbb{R} \mapsto \mathbb{R}^N$ is defined as
\begin{equation} \label{l1}
    \begin{aligned}
	\bigtriangleup_1(A,y,\epsilon) := \underset{z \in \mathbb{R}^N} {\text{argmin}} \ \lvert\lvert z \lvert\lvert_1  \ \ 
	                                   \text{s.t.} \ \  \lvert \lvert Az-y \lvert \lvert \leq \epsilon
\end{aligned}
\end{equation}
The $\ell_1$ minimization problem in (\ref{l1}) is a convex optimization problem and thus tractable. However, this computational tractability of $\ell_1$ minimization comes at the cost of increasing the number of measurements taken. For example if columns of $A$ are independent, identically distributed random vectors with any sub-Gaussian distribution, then $\bigtriangleup_1$ can recover any $k$-sparse vector $x$ when $n \gtrsim k\ log(\frac{N}{k})$ rather than the $n>2k$ property which is sufficient for recovery by $\bigtriangleup_0$.\newline
Several works have attempted to close the gap in the required number of measurements for recovery via $\ell_0$ and $\ell_1$ minimization problems, including solving a non-convex $\ell_p$ minimization problem with $0<p<1$ \cite{chartrand07letters,rayan:2010,Foucart08} and using prior knowledge about the signal \cite{weighted_l1:2011}. We will describe these in the next section. In this paper we propose to combine these approaches when there is prior information on the support of the signal. Specifically we introduce a weighted $\ell_p$ minimization algorithm and show that it outperforms both $\ell_p$ minimization and weighted $\ell_1$ minimization under certain circumstances.\newline
In Section 2, we briefly review various results on recovery by $\ell_1$, $\ell_p$, and weighted $\ell_1$ minimization. In Section 3, we describe the proposed recovery method based on weighted $\ell_p$ minimization, derive stability and robustness guarantees for this method and compare it with regular $\ell_p$ and weighted $\ell_1$. Specifically, we prove that the recovery guarantees for the weighted $\ell_p$ method with $0<p<1$ are better than those of weighted $\ell_1$ and regular $\ell_p$ when we have a prior support estimate with accuracy better than $50\%$. In Section 4, we explain the algorithmic issues that come with solving the proposed non-convex optimization problem and the approach we take to empirically overcome them. Next, we present numerical experiments where we apply the weighted $\ell_p$ method to recover sparse and compressible signals. In Section 5, we show the result of applying these algorithms to audio signals and seismic data. In Section 6, we provide the proof for our main theorem.
\section{Previous Work}
In this section, we state the recovery algorithms based on $\ell_p$ and weighted $\ell_1$ minimization, and the associated recovery guarantees. In both cases the restricted isometry constants play a central role.
\begin{dfn}
A matrix $A$ satisfies the restricted isometry property (RIP) of order $k$ with constant $\delta_k$ if for all $k$-sparse vectors $z \in \Sigma_k^N$,
\begin{equation} \label{irp}
 (1-\delta_k) \lvert\lvert z \lvert\lvert_2^2 \leq \lvert\lvert Az \lvert\lvert_2^2 \leq  (1+\delta_k) \lvert\lvert z \lvert\lvert_2^2.\newline
\end{equation}
\end{dfn}
\subsection*{Recovery by $\ell_p$ Minimization}
Chartrand\cite{chartrand07letters}, and Saab and Y{\i}lmaz \cite{rayan:2010}, cf. \cite{Foucart08}, considered the sparse recovery method based on $\ell_p$ minimization with $0<p<1$. Here, the $\ell_1$ norm in (\ref{l1}) is replaced by the $\ell_p$ quasi-norm. The decoder $\bigtriangleup_p: \mathbb{R}^{n\times N} \times \mathbb{R}^n \times \mathbb{R} \mapsto \mathbb{R}^N$ is defined as
\begin{equation} \label{lp}
    \begin{aligned}
	\bigtriangleup_p(A,y,\epsilon): =\underset{z \in \mathbb{R}^N} {\text{argmin}} \lvert\lvert z \lvert\lvert_p  \ \
	                                   \text{s.t.} \ \ \lvert \lvert Az-y \lvert \lvert \leq \epsilon .
\end{aligned}
\end{equation}
It was shown in \cite{chartrand07letters,rayan:2010,saab2008ssa,Foucart08} that recovery by $\ell_p$ minimization is stable and robust under weaker sufficient conditions than the analogous conditions for recovery by $\ell_1$ minimization. 
This result is made explicit by the following theorem from \cite{rayan:2010}. Note that setting $p=1$ below yields the robust recovery theorem of Cand{\'e}s, Romberg and Tao \cite{CRT05} with identical sufficient conditions and constants.
\begin{thm} \label{rayanlp}
(Saab and Y{\i}lmaz \cite{rayan:2010} ) Let $k$, $N$ be positive integers with $k<N$ and $p\in(0,1)$. Suppose that $x$ is an arbitrary vector in $\mathbb{R}^N$ and denote $x_{k}$ by the best $k$-term approximation of $x$. Let $y=Ax+e$ with $\|e\|_2 \leq \epsilon$. If $A$ satisfies  $\delta_{ak} + a^{\frac{2}{p}-1} \delta_{(a+1)k} <a^{\frac{2}{p}-1}-1$, for some $a\in \frac{1}{k} \mathbb{N}$, then
$$\lvert \lvert \bigtriangleup_p(A,y,\epsilon)-x \lvert \lvert_{2}^p \leq C_{1}^{\ell_p} \cdot \epsilon^p + C_{2}^{\ell_p} \cdot \frac{\lvert \lvert x-x_{k} \lvert \lvert_{p}^p}{k^{1-p/2}},$$
where $C_1^{\ell_p}$ and $C_2^{\ell_p}$ are given explicitly in \cite[Th. 2.1]{rayan:2010}.
\end{thm}
\begin{rem}
It is sufficient that $A$ satisfies
\begin{equation}\label{deltap}
\delta_{(a+1)k}<\hat{\delta}^{\ell_p} :=\frac{a^{\frac{2}{p}-1}-1}{a^{\frac{2}{p}-1}+1}
\end{equation}
for Theorem \ref{rayanlp} to hold (with same constants).
\end{rem}
\begin{rem}
Proposition 2.10 in \cite{rayan:2010} has compared the recovery guarantees of $\bigtriangleup_1$ and $\bigtriangleup_p$ in the noise free case. Assume there exists $k_1>1$ and $a\in \frac{1}{k_1} \mathbb{N}$ such that
$\delta_{(a+1)k_1} < \frac{a-1}{a+1}$. Then a standard result \cite[Theorem 1]{CRT05} guarantees that (\ref{l1}) can recover all $k_1$-sparse signals and Theorem \ref{rayanlp} guarantees that (\ref{lp}) can recover all $k_p$-sparse vectors where $k_p=\left\lfloor \frac{a+1}{a^{\frac{p}{2-p}}+1}k_1\right\rfloor$. Notice that $k_p>k_1$ when $p<1$.
\end{rem}
\subsection*{Recovery by Weighted $\ell_1$ Minimization}
The $\ell_1$ problem (\ref{l1}) does not use any prior information about the signal. In many applications it is possible to obtain a partially accurate estimate of the support---the set of indices of the large coefficients---of the signal. It was noted in \cite{weighted_l1:2011} that one can improve the recovery performance by incorporating the prior support information into the $\ell_1$-minimization-based recovery algorithm. In particular \cite{weighted_l1:2011} proposes the weighted $\ell_1$ decoder $\bigtriangleup_{1,w}:\mathbb{R}^{n\times N} \times \mathbb{R}^n \times \mathbb{R} \times \mathbb{R}^N \mapsto \mathbb{R}^N$ defined as
\begin{equation} \label{weightedl1}
\bigtriangleup_{1,w}(A,y,\epsilon,{\rm{w}}):= \underset{z \in \mathbb{R}^N} {\text{argmin}} \ \lvert\lvert z \lvert\lvert_{1,{\rm{w}}} \ \text{s.t.} \ \lvert \lvert Az-y \lvert \lvert \leq \epsilon,
\end{equation}
\normalsize
where ${\rm{w}} \in \lbrace\omega , 1\rbrace^N$ is the weight vector and $\| z \|_{1,{\rm{w}}}:=\Sigma_i {\rm{w}}_i \lvert \text{$z$}_i \lvert$ is the weighted $\ell_1$ norm of $z$. Given a support estimate $\widetilde{T} \subseteq \{1,...,N\}$ and assuming ${\rm{w}}_j=\omega <1$ for $j \in \widetilde{T}$ and $\rm{w}_j=1$ for $j\notin \widetilde{T}$, $\bigtriangleup_{1,w}$ enjoys better error bounds compared to $\bigtriangleup_{1}$ provided $\widetilde{T}$ is sufficiently accurate. The following theorem was proved in \cite{weighted_l1:2011}.
\begin{thm} \label{hassanl1}
(\cite{weighted_l1:2011} ) Let $x$ be an arbitrary vector in $\mathbb{R}^N$ and $y=Ax+e$ with $\|e\|_2 \leq \epsilon$. Denote $x_k$ by the best $k$-term approximation of $x$ with supp$\{x_k\}=T_0$. Let $\widetilde{T}$ be an arbitrary subset of $\{1,2,...,N\}$ and define $\rho$ and $\alpha$ such that $|\widetilde{T}|=\rho k$ and $|T_0 \cap \widetilde{T}|=\alpha\rho k$. 
Suppose there exists an $a\in \frac{1}{k}\mathbb{Z}$ with $a \geq (1-\alpha)\rho$ and $a>1$ and the measurement matrix $A$ has RIP with
$$\delta_{ak}+\frac{a}{(\omega+(1-\omega)\sqrt{1+\rho-2\alpha\rho})^2}\delta_{(a+1)k} < \frac{a}{(\omega+(1-\omega)\sqrt{1+\rho-2\alpha\rho})^2}-1$$
for some $ 0\leq \omega \leq 1 $. Then
$$\lvert \lvert \bigtriangleup_{1,w}(A,y,\epsilon,{\rm{w}})-x \lvert \lvert_{2} \leq  C_{1}^{w\ell_1} \epsilon + C_{2}^{w\ell_1} k^{\frac{-1}{2}}(\omega\|x-x_k\|_1 +(1-\omega)\|x_{\widetilde{T}^c\cap T_0^c}\|_1),$$
where $C_1^{w\ell_1}$ and $C_2^{w\ell_1}$ are given explicitly in \cite[Remark 3.1]{weighted_l1:2011}.
\end{thm}
\begin{rem}
It is sufficient that $A$ satisfies
\begin{equation}\label{deltawl1}
\delta_{(a+1)k}<\hat{\delta}^{w\ell_1} :=\frac{a-(\omega+(1-\omega)\sqrt{1+\rho-2\alpha\rho})^{2}}{a+(\omega+(1-\omega)\sqrt{1+\rho-2\alpha\rho})^{2}}
\end{equation}
\normalsize
for Theorem \ref{hassanl1} to hold (with same constants).
\end{rem}

\section{Main Results}
In this section we 
introduce the decoder $\bigtriangleup_{p,w}$ that is based on weighted $\ell_p$ minimization. For a given prior support estimate $\widetilde{T}$, $\bigtriangleup_{p,w}:\mathbb{R}^{n\times N} \times \mathbb{R}^n \times \mathbb{R} \times \mathbb{R}^N \mapsto \mathbb{R}^N$ is defined as
\begin{equation} \label{weightedlp}
\bigtriangleup_{p,w}(A,y,\epsilon,{\rm{w}}):=\underset{z \in \mathbb{R}^N} {\text{argmin}} \| z \|_{p,{\rm{w}}}\ 
	                                   \text{s.t.} \ \| Az-y \| \leq \epsilon \ with \ {\rm{w}}_i=\begin{cases} 1,&\text{if $i \in \widetilde{T}^c$}
\\
\omega ,&\text{if $i \in \widetilde{T}$}
\end{cases}.
\end{equation}
Here ${\rm{w}} \in \lbrace\omega , 1\rbrace^N$ is the weight vector and $\| z \|_{p,{\rm{w}}}:=(\Sigma_i {\rm{w}}_i^p \lvert \text{$z$}_i \lvert^p)^{\frac{1}{p}}$ is the weighted $\ell_p$ norm. Next we provide the stable and robust recovery conditions of this algorithm and compare it with weighted $\ell_1$ and $\ell_p$.
\subsection{Weighted $\ell_p$ Minimization with Estimated Support}
As mentioned in the previous section, one can improve the recovery guarantees of $\bigtriangleup_1$ by using $\bigtriangleup_p$ and by incorporating prior support information into the the optimization problem. In this section we provide the recovery conditions when we combine both these approaches. The following theorem states the main result.
\begin{thm}\label{maintheorem1}
Let $x$ be an arbitrary vector in $\mathbb{R}^N$ and $y=Ax+e$ with $\|e\|_2 \leq \epsilon$. Denote $x_k$ by the best $k$-term approximation of $x$ with supp$\{x_k\}=T_0$. Let $\widetilde{T}$ be an arbitrary subset of $\{1,2,...,N\}$ and define $\rho$ and $\alpha$ such that $|\widetilde{T}|=\rho k$ and $|T_0 \cap \widetilde{T}|=\alpha\rho k$. Suppose there exist an $a\in \frac{1}{k}\mathbb{Z}$, with $a \geq (1-\alpha)\rho$ and $a>1$ and the measurement matrix $A$ has RIP with
\footnotesize
\small
\normalsize
$$\delta_{ak}+\frac{a^{\frac{2}{p}-1}}{(\omega^p+(1-\omega^p)(1+\rho-2\alpha\rho)^{1-\frac{p}{2}})^{\frac{2}{p}}}\delta_{(a+1)k} < \\ \frac{a^{\frac{2}{p}-1}}{(\omega^p+(1-\omega^p)(1+\rho-2\alpha\rho)^{1-\frac{p}{2}})^{\frac{2}{p}}}-1,
$$
\normalsize
for some $ 0\leq \omega \leq 1 $ and $0<p<1$. Then
\begin{equation}\label{perror}
\|\bigtriangleup_{p,w}(A,y,\epsilon,{\rm{w}})-x \|_{2}^p \leq C_{1}  \epsilon^p + C_{2} k^{\frac{p}{2}-1}(\omega^p\|x-x_k\|_p^p +(1-\omega^p)\|x_{\widetilde{T}^c\cap T_0^c}\|_p^p).
\end{equation}
\end{thm}
\begin{rem}
Note that 
$\rho$ denotes the ratio of the size of the estimated support to the size of the actual support of $x_k$ and $\alpha$ denotes the accuracy of our estimate which is the ratio of the size of $\widetilde{T} \cap T_0$, to the the size of our estimate $\widetilde{T}$.
\end{rem}
\begin{rem}
The constants $C_1$ and $C_2$ are explicitly given in (\ref{C1C2}) in Section 6.
\end{rem}

\begin{rem}
It is sufficient that $A$ satisfies
\begin{equation}\label{delta2p}
\delta_{(a+1)k}<\hat{\delta}^{w\ell_p} :=\frac{a^{\frac{2}{p}-1}-(\omega^p+(1-\omega^p)(1+\rho-2\alpha\rho)^{1-\frac{p}{2}})^{\frac{2}{p}}}{a^{\frac{2}{p}-1}+(\omega^p+(1-\omega^p)(1+\rho-2\alpha\rho)^{1-\frac{p}{2}})^{\frac{2}{p}}}
\end{equation}
\normalsize
for Theorem \ref{maintheorem1} to hold, i.e., to guarantee stable and robust recovery described in the theorem with same constants $C_1$ and $C_2$. Setting $\omega=1$ gives us the the sufficient conditions for recovery by $\bigtriangleup_p$ and setting $p=1$ derives the sufficient recovery conditions for recovery by $\bigtriangleup_{1,w}$. Notice that these conditions are in terms of bounds on RIP constants. In the remainder of this section we compare these bounds.
\end{rem}
\subsection{Comparison with Weighted $\ell_1$ Recovery}
\begin{figure}[t]
\centering
\subfloat[][$\hat{\delta}^{w \ell_p}$ vs $\omega$]{
\includegraphics[width=0.8\textwidth]{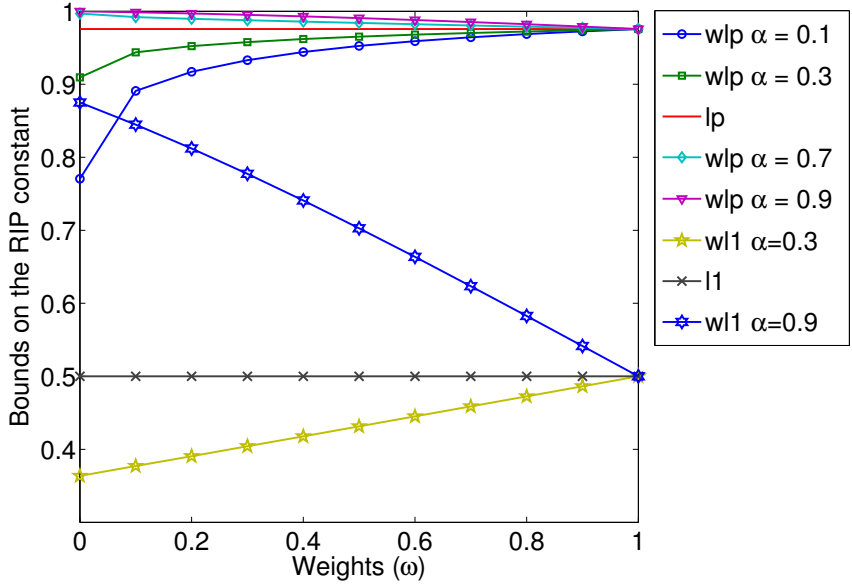}
\label{fig:subfig1}}
\caption{Comparison of the sufficient conditions for recovery with weighted $\ell_p$ reconstruction with various $\alpha$. In all figures, we set a = 3 and $\rho=1$ and $p=\frac{2}{5}$.}
\label{fig1paper}
\end{figure}

\begin{figure} [ht]
\centering
\includegraphics[height=70 mm]{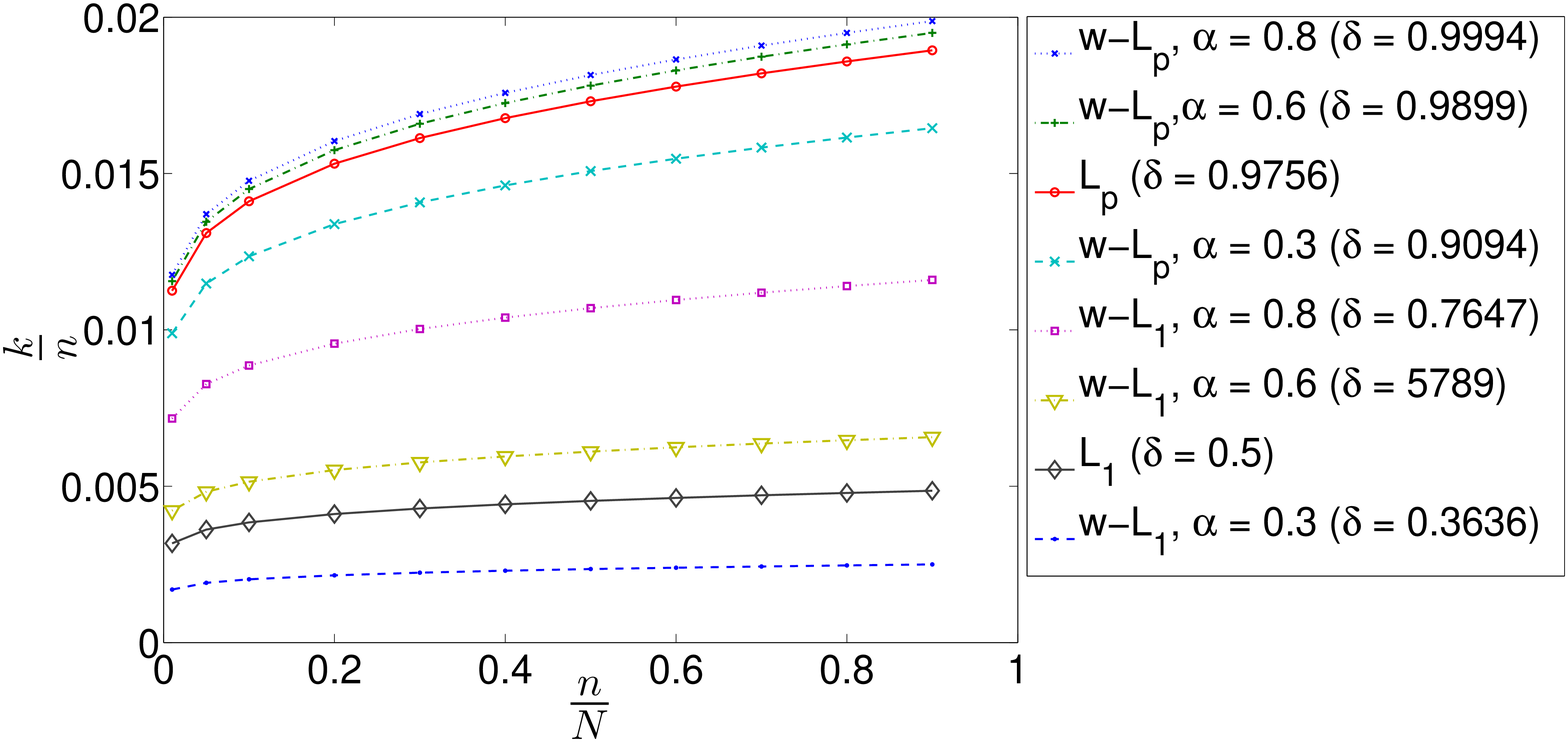}
\caption{Comparison between phase diagrams of measurement matrices with Gaussian entries satisfying the sufficient recovery conditions of weighted $\ell_p$ and weighted $\ell_1$ minimization with $\omega=0$. Points below each curve determine the sparsity-undersampling ratios that satisfy the sufficient bounds on the RIP constants introduced in (\ref{deltawl1}) and (\ref{delta2p}).}
\label{fig2paper}
\end{figure}

In this section we compare the conditions for which Theorem \ref{maintheorem1} holds with the corresponding conditions of Theorem \ref{hassanl1}. Following observation is easy to verify.
\begin{per}
Let $C_1$, $C_2$, $C_1^{w\ell_1}$ and $C_2^{w\ell_1}$ be as defined above. If $p=1$ then $C_1=C_1^{w\ell_1}$ and $C_2=C_2^{w\ell_1}$ and the sufficient condition for Theorem \ref{maintheorem1} would be identical to Theorem \ref{hassanl1}.
\end{per}
Figure \ref{fig1paper} illustrates how the sufficient conditions on the RIP
constants vary with $\alpha$ and $\omega$ in the case of weighted $\ell_1$ and weighted $\ell_p$. In particular these sufficient conditions are introduced in Theorem \ref{hassanl1} and Theorem \ref{maintheorem1}, i.e., $\hat{\delta}^{w\ell_1}$ defined in (\ref{deltawl1}) and $\hat{\delta}^{w\ell_p}$ defined in (\ref{delta2p}) which determine bounds on the RIP constants. Here we plot $\hat{\delta}^{w\ell_p}$ versus $\omega$ for weighted $\ell_1$ ($p=1$) and weighted $\ell_p$ ($0<p<1$) with different values of $\alpha$ when $a=3$ and $p=\frac{2}{5}$. The bounds on RIP constants gets larger as $\alpha$ increases. 
Note that when $\alpha=0.5$ the sufficient conditions for recovery by weighted $\ell_p$ would be identical to sufficient conditions for recovery by standard $\ell_p$ for $0<p<1$. Comparing these results with recovery by weighted $\ell_1$, we see that in recovery by weighted $\ell_p$ the measurement matrix $A$ has to satisfy much weaker conditions than the analogous conditions in recovery by weighted $\ell_1$ even when we do not have a good support estimate.
It is worth comparing the sufficient recovery conditions for the special case of zero weight. As seen in Figure \ref{fig1paper} setting $\omega=0$ is beneficial when $\alpha >0.5$. 
Figure \ref{fig2paper} compares the recovery guarantees we obtain in the zero-weight case for weighted $\ell_p$ and weighted $\ell_1$ minimization. Specifically, we present the phase diagrams of measurement matrices $A$ with Gaussian entries that satisfy the conditions on the restricted isometry constants 
$\delta_{(a+1)k}$ given in (\ref{deltawl1}) and (\ref{delta2p}) with $\omega =0$, $\rho = 1$, and $\alpha = 0.3$, $0.6,$ and $0.8$. Phase diagrams are calculated using the upper bounds on the RIP constants derived in \cite{Bah_Tanner_RIPbounds:2010} and reflect the sparsity levels for which the theorems guarantee exact signal recovery as a function of the aspect ratio of the measurement matrix $A$.
\subsection{Comparison with $\ell_p$ Recovery}
\begin{figure}[h]
\centering
\subfloat[$C_1$ vs $\omega$]{
\includegraphics[width=0.5\textwidth, height=0.4\textwidth]{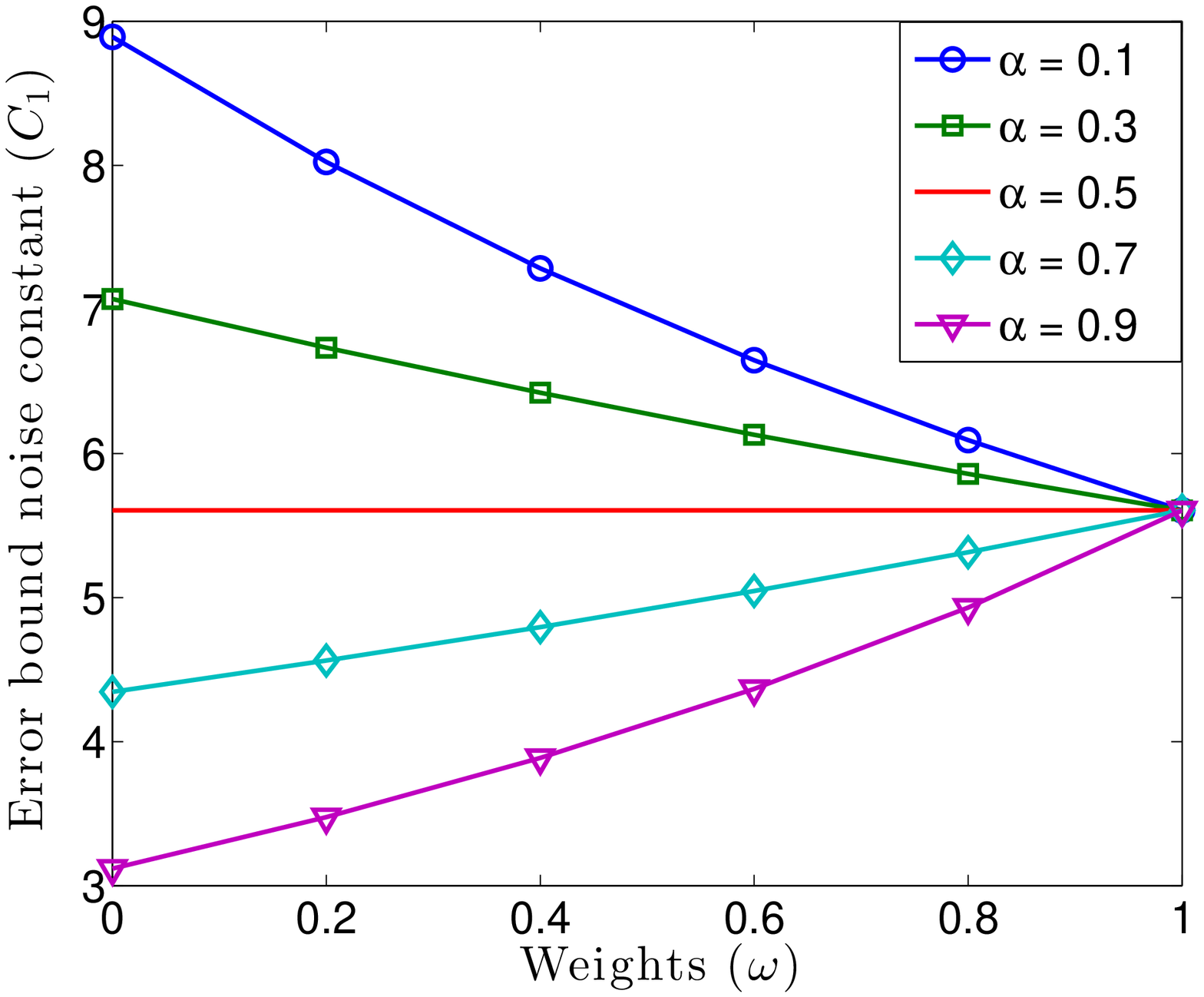}
\label{fig:subfig2}}
\subfloat[$C_2$ vs $\omega$]{
\includegraphics[width=0.53\textwidth, height=0.42\textwidth]{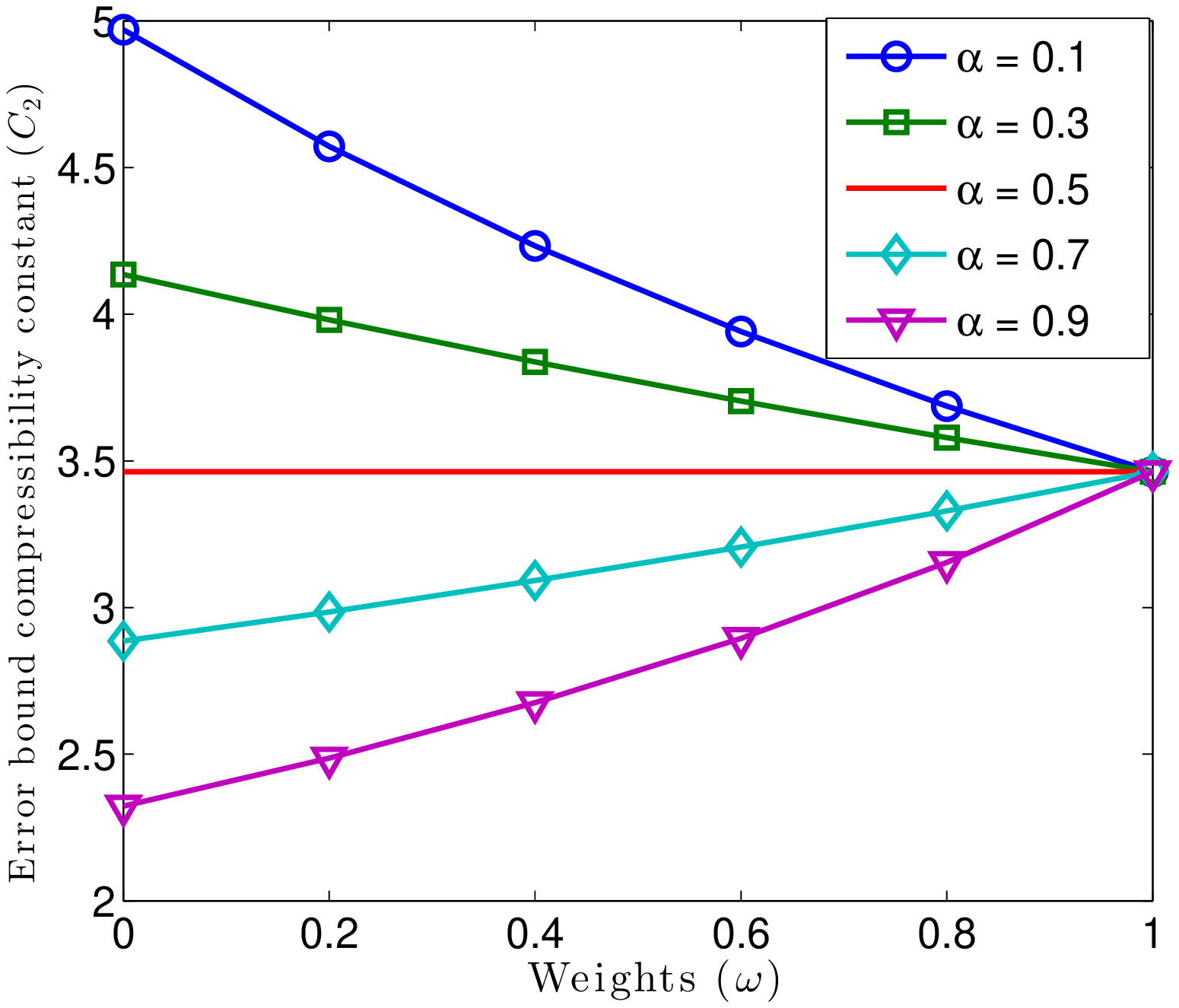}
\label{fig:subfig3}}
\caption{Comparison of the recovery constants for weighted $\ell_p$ reconstruction with various $\alpha$. In all the Figures, we set a = 3 and $\rho=1$ and $p=\frac{2}{5}$ .}
\label{constants}
\end{figure}

In this section we compare the sufficient conditions of Theorem \ref{rayanlp} and Theorem \ref{maintheorem1}. The following is easy to check.
\begin{per}\label{prop12}
Let $C_1$, $C_2$, $C_1^{\ell_p}$ and $C_2^{\ell_p}$ be as defined above .\newline
(i) If $\alpha=0.5$ then again $C_1=C_1^{\ell_p}$ and $C_2=C_2^{\ell_p}$ and the sufficient condition for Theorem \ref{maintheorem1} would be identical to Theorem \ref{rayanlp}.\newline
(ii) Suppose $0 \leq \omega <1$. Then $C_1<C_1^{\ell_p}$ and $C_2<C_2^{\ell_p}$ if and only if $\alpha>0.5$
\end{per}
Proposition \ref{prop12} reflects the results shown in Figure \ref{constants}. Figures \ref{constants}.a and \ref{constants}.b show how constants $C_1$ and $C_2$ in (\ref{perror}) change with $\omega$ for different values of $\alpha$. Notice that constants decrease when we increase $\alpha$.\newline
When $\alpha<0.5$, i.e., when our estimate is less than $50\%$ accurate, using bigger  weights results in more robust recovery, which is useful when the accuracy of the estimate is not guaranteed to be high. 
For all values of $\omega < 1$, having a support estimate accuracy $\alpha > 0.5$ results in a weaker condition on the RIP constant and smaller error bound constants compared with the conditions of standard $\ell_p$. On the other hand, if $\alpha < 0.5$, i.e., the support estimate has low accuracy, then standard $\ell_p$ has weaker sufficient recovery conditions and smaller error bound constants compared to weighted $\ell_p$. This behaviour is similar to that derived for weighted $\ell_1$ minimization in \cite{weighted_l1:2011}.
\section{Numerical Experiments}
\subsection{Algorithmic Issues}
Before we present numerical experiments, we describe the algorithm that we used to approximate $\bigtriangleup_{p,w}$, i.e., to \enquote{solve} the weighted $\ell_p$ minimization problem.\newline
To this day, there is no algorithm that provably solves this non-convex optimization problem. On the other hand, 
there are a few algorithms which are commonly used to attempt to solve this minimization problem. These include simple modifications of well-known algorithms such as the projected gradient method \cite{chartrand07letters}, the iterative reweighted $\ell_1$ method \cite{irlp}, and the iterative reweighted least squares method \cite{chartrand-2008-iteratively}. Since the $\ell_p$ minimization problem is non-convex and several local minima exist, these algorithms attempt to converge to local minima that are close to the global minimizer of the problem. To that end, the only proofs of global convergence that currently exist assume that the global minimizer can be found if a feasible point can be found. However, numerical experiments show that these algorithms perform well, for example, when the measurement matrix has i.i.d. Gaussian random entries. To produce the numerical experiments below, we have used the projected gradient method which is described next.\newline
The algorithm starts by minimizing a smoothed $\ell_p$ objective given by $(\sum_i (x_i^2 + \sigma)^{p/2})^{1/p}$ instead of of the $\ell_p$ norm. The smoothing parameter $\sigma$ is initialized with a large value of 10. The algorithm follows by taking a projected gradient step and reducing the value of $\sigma$. In every iteration, the new iterant is projected onto the affine space $Ax = b$.
Algorithm \ref{alg:pgm} explains the details of this algorithm. Here $ \nabla (f_x)_i$ = $ p \times {\rm{w}}_i^p \times (x^{(t)}_i \times (x^{(t)}_i)^{*}+\sigma^2)^{p/2-1} \times x^{(t)}_i$.
\begin{algorithm}\caption{Modified projected gradient method}
\begin{algorithmic}[1]\label{alg:pgm}
\STATE \textbf{Input} $b = Ax + e$, $p$, $A$, $\omega_i \in [0,1]$ for all $i \in {1...N}$
\STATE \textbf{Output} $x^{(t)}$
\STATE \textbf{Initialize} $\sigma =10$, $t = 0$, $x^{(0)} = A^H b$, $[M \ \ N]=size(A)$, $Q=A^{\dagger} \times A$, ${\small
\mathrm{w}_i = \left\{
\begin{array}{l}
	\omega, \quad i \in \Lambda \\
	1, \quad i \in \Lambda^c \\
\end{array} \right.}
$   
\LOOP
\STATE $f_x=\sum_i (\rm{w}_i^2 \times (x^{(t)^2}_i + \sigma))^{p/2}$
\STATE $d=- \nabla (f_x)$
\STATE $pd=d-Q \times d$
\STATE $t = t + 1$
\STATE line search
\STATE $x^{(t)}=x^{(t-1)}+l \times pd$
\STATE Indicator=$\frac{\sqrt{1-p} \times x^{(t)}}{1-\sqrt{p}}$
\STATE Idx=find(Indicator < $\mathrm{w} \times \sigma$)
\STATE $\sigma = \min( 0.98 \times \sigma, \max(\textrm{Indicator}))$
\ENDLOOP
\end{algorithmic}
\end{algorithm}\vspace{-0.0in}

Next, we provide numerical results to show how $\bigtriangleup_{p,w}$ improves the recovery conditions of sparse and approximately sparse signals compared to $\bigtriangleup_p$ and $\bigtriangleup_{1,w}$. We show the results for sparse and compressible signals where we use Algorithms \ref{alg:pgm} to solve the weighted $\ell_p$ minimization problem.
\subsection{Numerical Experiments: The Sparse Case}
\begin{figure*}[!t]
\centering
\subfloat[ No-noise]{
\includegraphics[width=164 mm , height=62 mm]{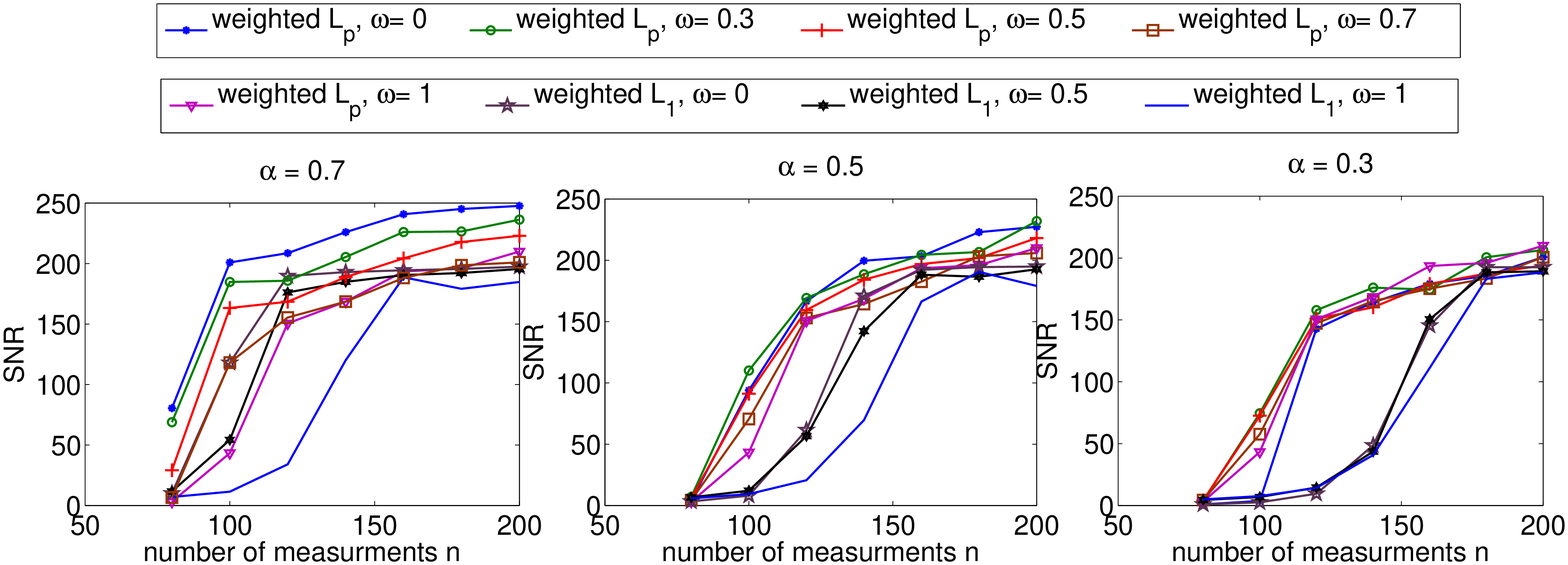}
\label{fig:subfig1}}\newline
\subfloat[ $5\%$ noise]{
\includegraphics[width=160 mm , height=49 mm]{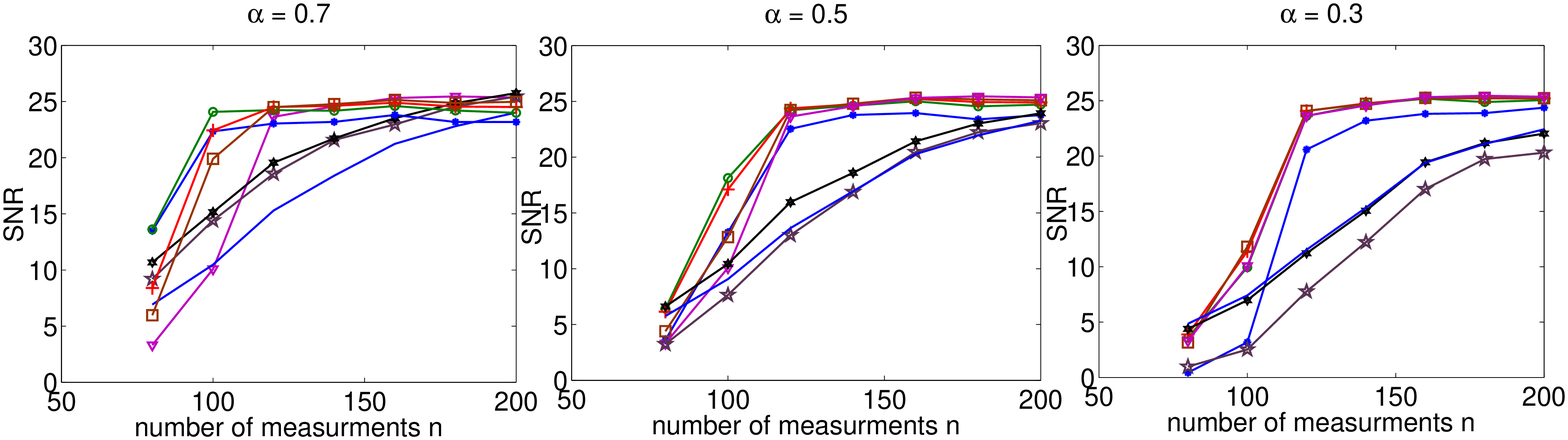}
\label{fig:subfig4}}
\caption{Comparison of Performance of weighted $\ell_p$ and weighted $\ell_1$ recovery in terms of SNR averaged over 10 experiments for sparse signals with variable weights and measurements and $\rho=1$ and $p=0.5$.}
\label{fig3paper}
\end{figure*}

In this section, we compare the performance of $\bigtriangleup_{1,w}$ in recovering exactly sparse signals for various values of $p$ and weight ${\rm{w}}$ including $p=1$, which corresponds to weighted $\ell_1$ of \cite{weighted_l1:2011} and ${\rm{w}}=[1,\ 1,\ \dots,\ 1 ]^{\text{T}}$, which corresponds to $\ell_p$ minimization. Specifically, we create $40$-sparse signals $x \in \mathbb{R}^{500}$, and obtain (noisy) compressed measurements of $x$ via $y=Ax+e$ where $A$ is chosen to be an $n \times 500$ Gaussian matrix with $n$ varying between 80 and 200. In the case of noisy measurements, $e$ is drawn from uniform distribution on the sphere and normalized such that $\frac{\| e \|_2}{\| x \|_2}=0.05$. Figure \ref{fig3paper} shows the reconstruction signal-to-noise ratio (SNR)
averaged over 10 experiments as a function of the number of the measurements obtained using weighted $\ell_p$ and weighted $\ell_1$ minimization. Figures \ref{fig3paper}.a and \ref{fig3paper}.b show the noise-free case and the noisy case, respectively. In both scenarios, we try different levels of prior support estimate accuracy $\alpha$, i.e., $\alpha\in \{0.3, 0.5, 0.7\}$ with weighted $\ell_p$ ($p=0.5$) and weighted $\ell_1$. Here the SNR is measured in dB and is given by
\begin{equation}\label{SNR}
\text{SNR}(x,\hat{x})=10 \log_{10}(\frac{\|x\|_2^2}{\|x-\hat{x}\|_2^2}).
\end{equation}
Figure \ref{fig3paper}.a illustrates that, in the noise free case, the experimental results are consistent with the theoretical results derived in Theorem \ref{maintheorem1}. More precisely, when $\alpha>0.5$ the best recovery is achieved when the weights are set to zero and as $\alpha$ decreases, the best recovery is achieved when larger weights are used. Also weighted $\ell_p$ is recovering significantly better than weighted $\ell_1$, especially when we have few measurements, which is consistent with our analysis in Section 3.
\begin{rem}
In Figures \ref{fig1paper} and \ref{constants} we can see that when $\alpha$<0.5 both the sufficient recovery conditions and error bound constants point towards using $\omega = 1$. However, Figure \ref{fig3paper} suggests that this is not always true. We attribute this behavior to the best $k$-term approximation term in the error bound of Theorem \ref{maintheorem1}. Consider the noise free case where the error bound becomes $\| x^*-x \|_{2}^p \leq C_{2} k^{\frac{p}{2}-1}(\omega^p\|x-x_k\|_p^p +(1-\omega^p)\|x_{\widetilde{T}^c\cap T_0^c}\|_p^p)$. Notice that on $T_0^c$, $x_k=0$ so we have $\|x_{\widetilde{T}^c\cap T_0^c}\|=\|(x-x_k)_{\widetilde{T}^c\cap T_0^c}\|$ which means that $\|x_{\widetilde{T}^c\cap T_0^c}\|_p^p \leq \|x-x_k\|_p^p$. Therefore, increasing $\omega$ increases $\omega^p\|x-x_k\|_p^p +(1-\omega^p)\|x_{\widetilde{T}^c\cap T_0^c}\|_p^p$. On the other hand, as we can see in Figure \ref{constants}, the constant $C_2$ decreases as $\omega$ increases. Consequently when the algorithm cannot recover the full support of $x$, i.e., when $\|x-x_k\|>0$, an intermediate value of $\omega$ in $(0,1)$ may result in the smallest recovery error. A full mathematical analysis of the above observations needs to take into account all the interdependencies between $\omega, k, \alpha$ and the parameters in Theorem \ref{maintheorem1} which is beyond the scope of this paper.\newline
Figure \ref{fig3paper}.b shows results for the noisy case. Using intermediate weights results in best recovery and weighted $\ell_p$ is outperforming weighted $\ell_1$ especially when we have few measurements.
\end{rem}
\subsection{Numerical Experiments: The Compressible Case}
\begin{figure*}[!t]
\centering
\subfloat[ No-noise]{
\includegraphics[width=164 mm , height=67 mm]{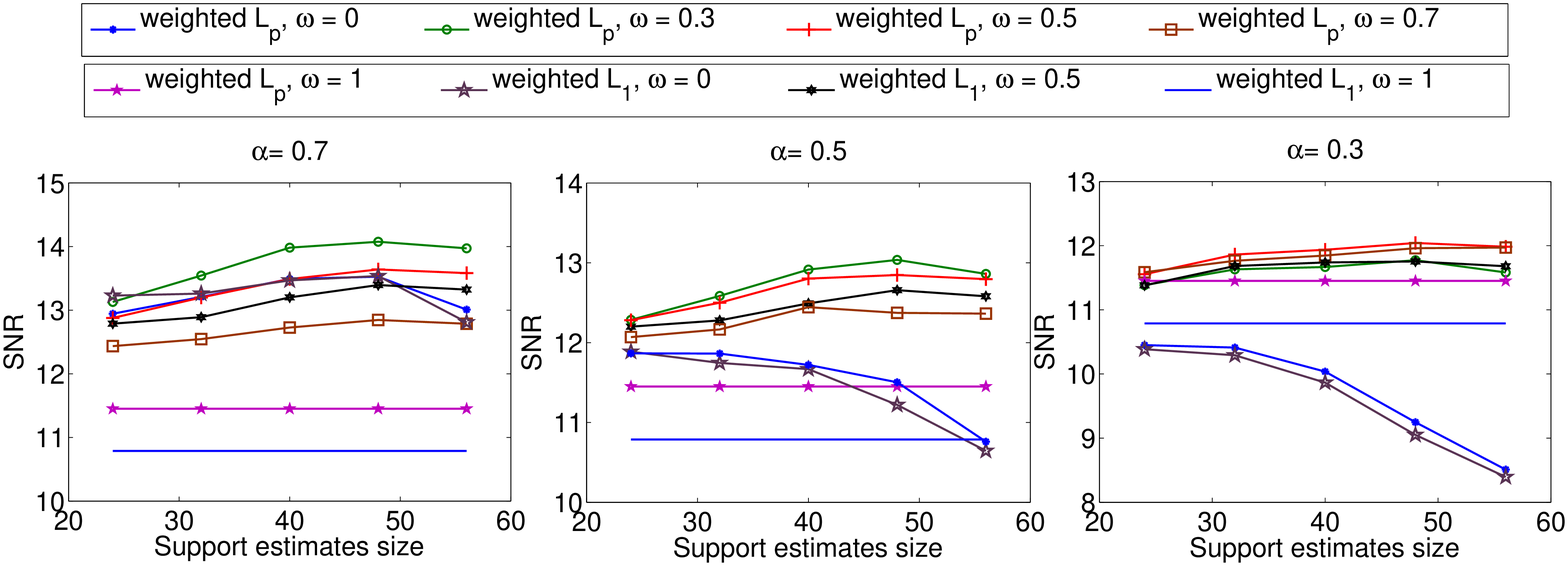}
\label{fig:subfig1}}\newline
\subfloat[ $5\%$ noise]{
\includegraphics[width=164 mm , height=51 mm]{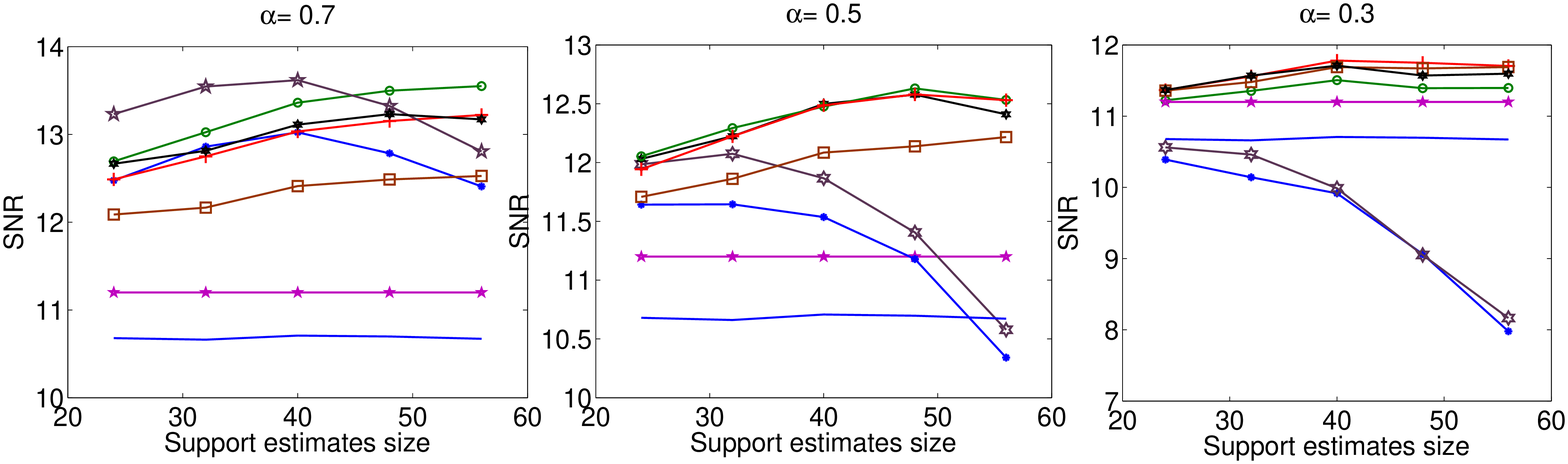}
\label{fig:subfig4}}
\caption{Comparison of performance of weighted $\ell_p$ and weighted $\ell_1$ recovery in terms of SNR averaged over 20 experiments for compressible signals $x$ with $n = 100, N = 500$. The coefficients decay with a power $d= 1.1$. The accuracy of the support estimate $\alpha$ is calculated with respect to the best $k = 40$ term approximation.}
\label{fig6paper}
\end{figure*}

In this section we consider signals $x \in \mathbb{R}^{500}$ such that $x_j=j^{-d}$ for some $d>1$.  Figure \ref{fig6paper} shows the average SNR over 20 experiments---20 Gaussian measurement matrices $A$ with the same signal $x$---when $n=100$ and $d=1.1$. We generate support estimates that target to find the locations of the largest 40 entries of $x$, i.e., a support estimate with accuracy $\alpha=1$ and relative size $\rho=1$ is $\{1,\dots,40\}$. Figure \ref{fig6paper}.a shows the no-noise case and Figure \ref{fig6paper}.b has $5\%$ noise.
As we can see 
using intermediate weights results in better reconstruction. 
When the measurements are noisy, unlike the sparse case, using weighted $\ell_p$ for recovering compressible signals doesn't give us much better results than weighted $\ell_1$, specifically in Figure \ref{fig6paper}.b when $\alpha=0.7$ we see that weighted $\ell_1$ with zero weight is recovering better than weighted $\ell_p$. We believe that this is a result of the algorithm we are using. As we said before we don't have any proof for global convergence of the algorithm and the projected gradient algorithm handles the local minima by a smoothing parameter $\sigma$. In the noisy compressible case we have lots of these local minimums which may be a reason that in some of the compressible noisy cases we see that weighted $\ell_1$ is recovering better than weighted $\ell_p$.\newpage

\section{Stylized Applications}
In this section, we apply standard and weighted $\ell_p$ minimization to recover real audio and seismic signals
that are compressively sampled.
\subsection{Audio Signals}
\begin{figure}[ht]
\centering
\includegraphics[width=0.6\textwidth]{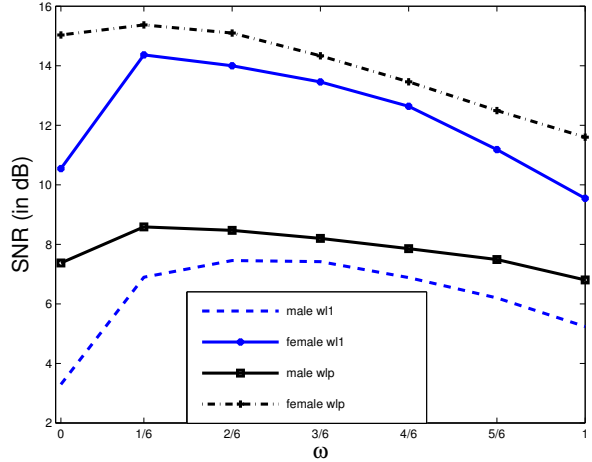}
\label{fig:subfig1}
\caption{SNRs of reconstructed audio signals from compressed sensing measurements plotted against $\omega$ via weighted $\ell_1$ weighted $\ell_p$ with $p=\frac{1}{2}$. An intermediate value of $\omega$ yields the best performance.}
\label{fig8paper}
\end{figure}

In this section we examine the performance of weighted $\ell_p$ minimization for the recovery of compressed sensing measurements of speech signals. Here the speech signals are sampled at 44.1 kHz and we randomly choose only $\frac{1}{4}$th of the samples. 
Assuming that $s$ is the speech signal, we obtain the measurements $y = Rs$ where $R$ is a restriction of the identity operator.\newline
We divide our measurements $y$ into 21 blocks, i.e., $y=[y_1^T,y_2^T,...]^T$. Assuming the speech signal is compressible in DCT domain
, we try to recover it using each block measurement.\newline
Doing this reduces the size of the problem and considering the fact that the support set corresponding to the largest coefficients doesn't change much from one block to another, we can use the indices of the largest coefficients of each block as a support estimate for the next one. For each block, we find the speech signal by solving $y_j=R_js_j$, where $R_j\in \mathbb{R}^{n_j\times N}$ is the associated restriction matrix. We also know that speech signals have large low-frequency coefficients, so we use this fact and the recovered signal at previous block to build our support estimate and find the speech signal at each block by weighted $\ell_p$ minimization. We choose the support estimate to be $\widetilde{T}=\widetilde{T}^1 \cup \widetilde{T}^2$, where $\widetilde{T}^1$ is the set corresponding to frequencies up to 4 kHz and $\widetilde{T}^2$ is the set corresponding to the largest $\frac{n_j}{16}$ recovered coefficients of the previous block---for the first block $\widetilde{T}^2$ is empty. The results of using weighted $\ell_p$ and weighted $\ell_1$ for reconstruction two audio signals---one male and one female---are illustrated in Figure \ref{fig8paper}. Here $N = 2048$ , and $\omega \in{0,\frac{1}{6}},\frac{2}{6},..,1$. Weighted $\ell_p$ gives about 1-dB improvement in reconstruction.
\subsection{Seismic Signals}
\begin{figure*}[!t]
\centering
\subfloat[][]{
\includegraphics[width=0.35\textwidth]{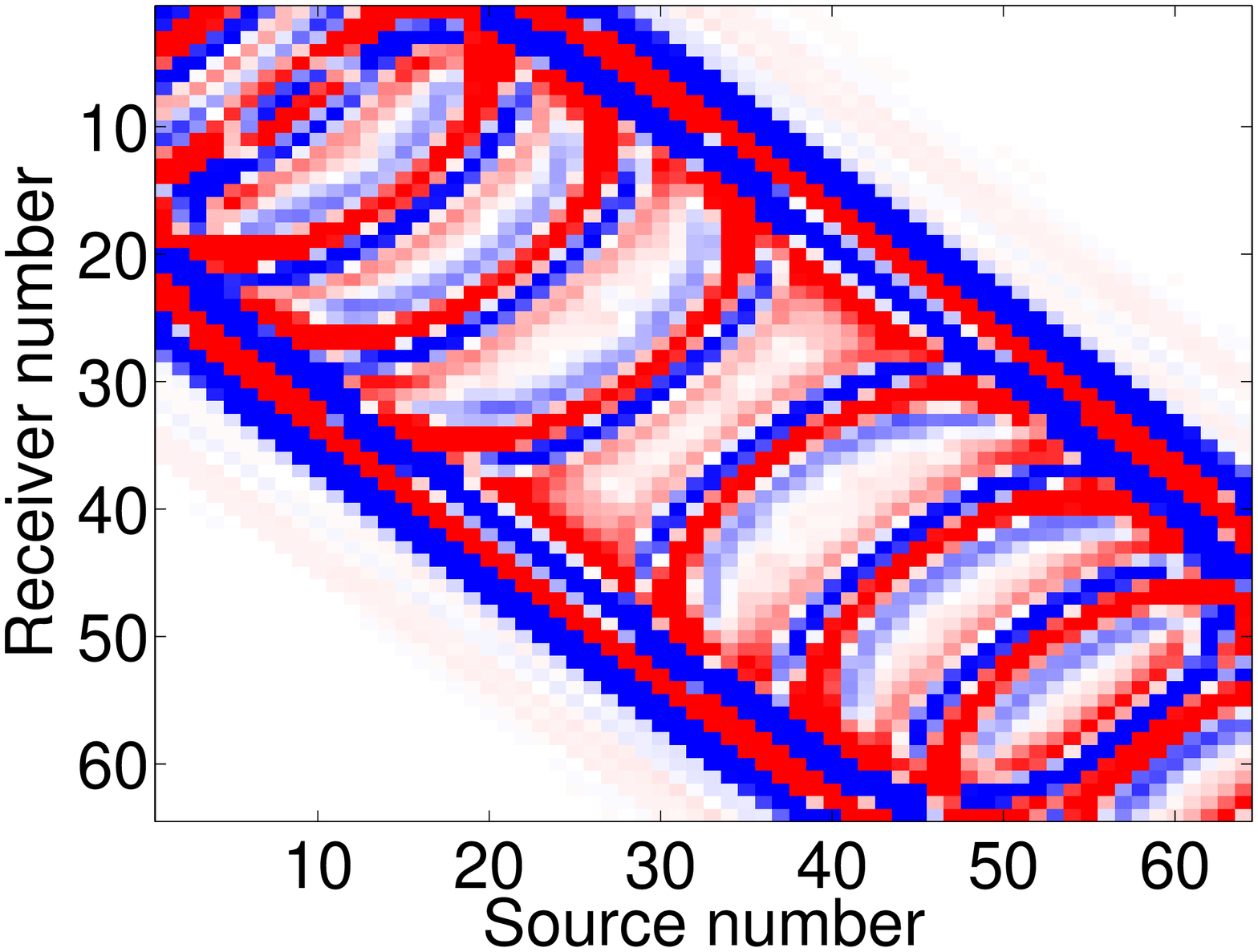}
\label{fig:subfig1}}
\subfloat[][]{
\includegraphics[width=0.35\textwidth]{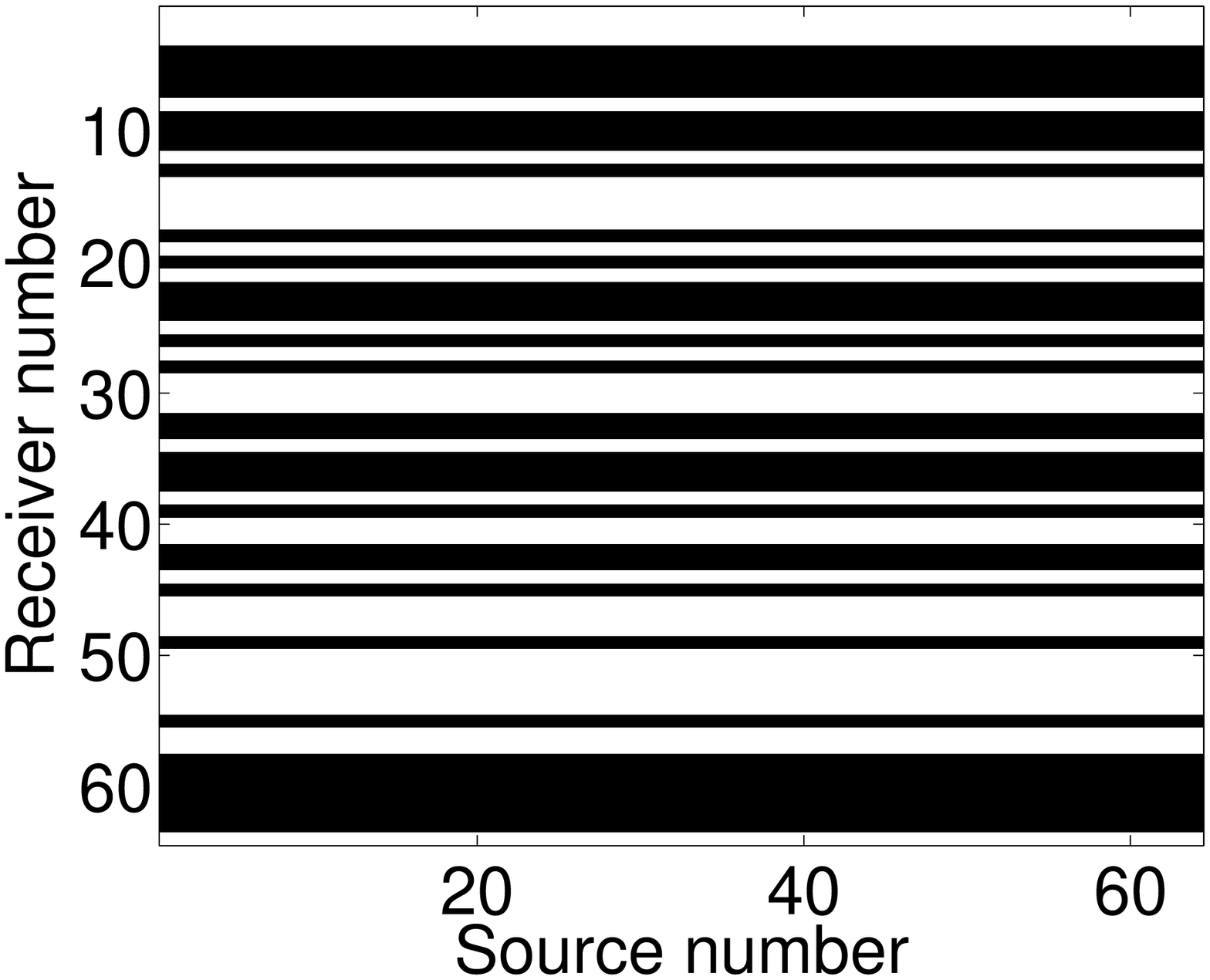}
\label{fig:subfig2}}
\subfloat[][]{
\includegraphics[width=0.35\textwidth]{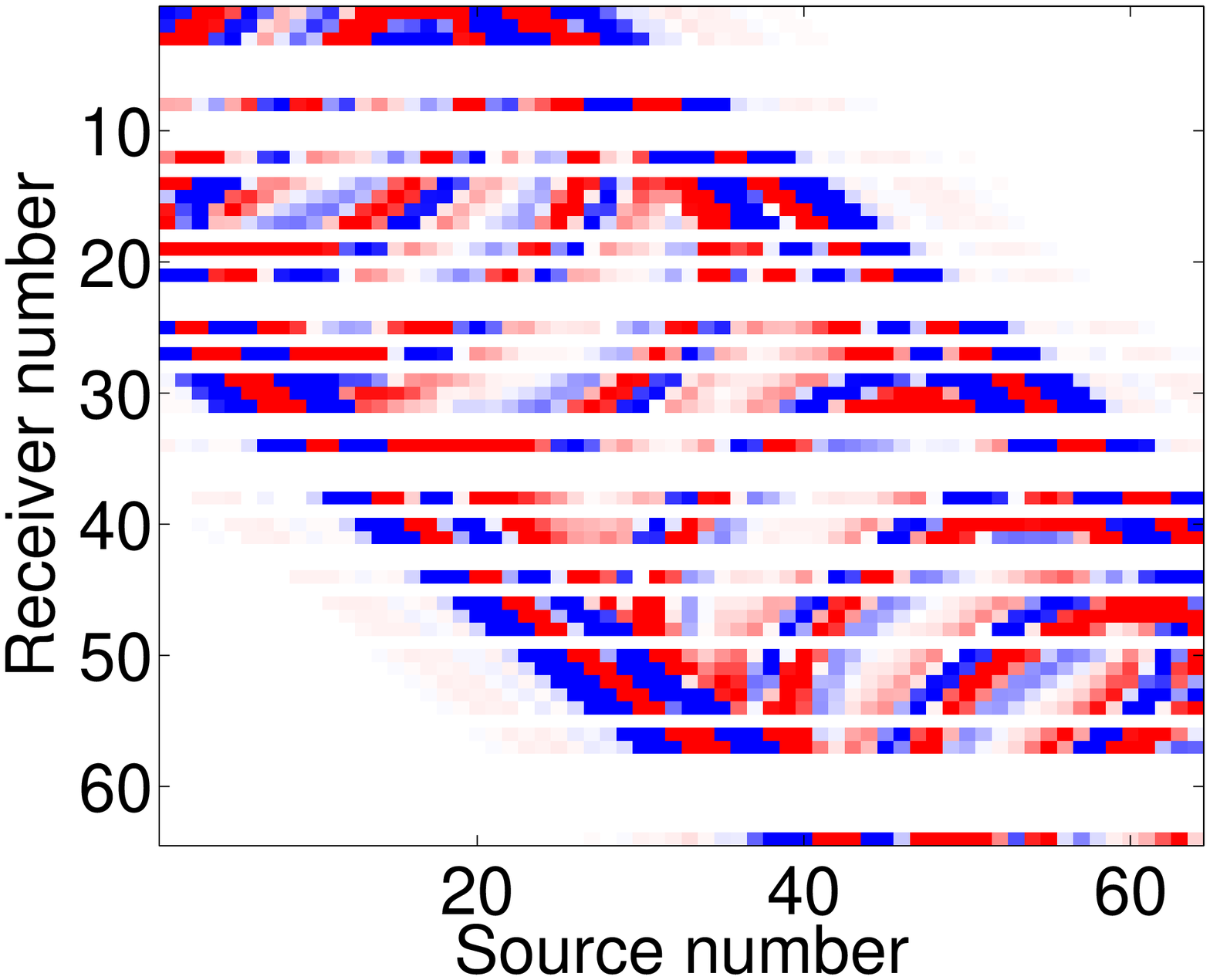}
\label{fig:subfig3}}
\caption{(a) Example of a high resolution time slice at $t =0.32$s in the source-receiver domain, (b) the random subsampling mask where the black lines correspond to the locations of inactive receivers, and (c) the subsampled time slice. Subsampling ratio is 50$\%$.}
\label{mask}
\end{figure*}

The problem of interpolating irregularly sampled and incomplete seismic data to a regular periodic grid often occurs in 2D and 3D seismic settings \cite{hassanfelix}. 
Assume that we have $N_s$ sources located on earth surface which send sound waves into the earth and $N_r$ receivers record the reflection in $N_t$ time samples. Hence the seismic data is organized in a 3-D seismic line with $N_s$ sources, $N_r$ receivers, and $N_t$ time samples. Rearranging the seismic line, we have a signal $f \in \mathbb{R}^N$, where $N=N_s N_r N_t$. Assume $x=Sf$ where $x$ is the sparse representation of $f$ in curvelet domain. We want to recover a very high dimensional seismic data volume $f=S^{*}x$ by interpolating between a smaller number of measurements $b=RMS^{*}x$, where $R$ is a restriction matrix, $M$ represents the basis in which the measurements are taken, and $S$ is the 2D curvelet transform. Seismic data is approximately sparse in curvelet domain and hence the interpolation problem becomes that of finding the curvelet synthesis coefficients with the smallest $\ell_1$ norm that best fits the randomly subsampled data in the physical domain \cite{CandesDemanet2005,Herrmann_curvelet2008}. We partition the seismic data volume into frequency slices and approximate $x^{(1)}$ by $\widetilde{x}^{(1)} := \bigtriangleup_{p}(R^{(1)}MS^*,b^{(1)},\epsilon)$ where $\epsilon$ is a small number (estimate of the noise level) and $R^{(1)}$ is the subsampling operator restricted to the first partition and $b^{(1)}$ is the subsampled measurements of the data $f^{(1)}$ in the first partition. After this we use the support of each recovered partition as a support estimate for next partition. In particular for $j \geq 1$ we approximate $x^{(j+1)}$ by $\widetilde{x}^{(j+1)} := \bigtriangleup_{p,{\rm{w}}}(R^{(j)}MS^{H},b^{(j)},\epsilon,{\rm{w}})$ where ${\rm{w}}$ is the weight vector which puts smaller weights on the coefficients that correspond to the support of the previous recovered partition. In \cite{hassanfelix} the performance of weighted $\ell_1$ minimization has been tested for recovering a seismic line using 50\% randomly subsampled receivers. Exploiting the ideas in \cite{hassanfelix} we test the weighted $\ell_p$ minimization algorithm to recover a test seismic problem when we subsample 50\% of the the receivers using the mask shown in Figure \ref{mask}.b. We 
omit the details of this algorithm as it mimics the steps taken in \cite{hassanfelix} when weighted $\ell_1$ is replaced by weighted
$\ell_p$.\newline
The seismic line at full resolution has $N_s = 64$ sources, $N_r =64$ 
receivers with a sample distance of 12.5 meters, and $N_t = 256$ time samples acquired with a sampling interval of 4 milliseconds. Consequently, it contains samples collected in a 1s temporal window with a maximum frequency of 125 Hz. To access frequency slices, we take the one dimensional discrete Fourier transform (DFT) of the data along the time axis. We solve the $\ell_p$ and weighted $\ell_p$ minimization problems. In the $j+1$-th partition, the support estimate set is derived from the largest analysis coefficients $SS^{H}\widetilde{x}^{(j)}$ of the previously recovered partition. Moreover, $p$ is set to be $0.5$ and the weight is set to 0.3.\newline
\begin{figure*}[!t]
\centering
\subfloat[][]{
\includegraphics[width=0.48\textwidth]{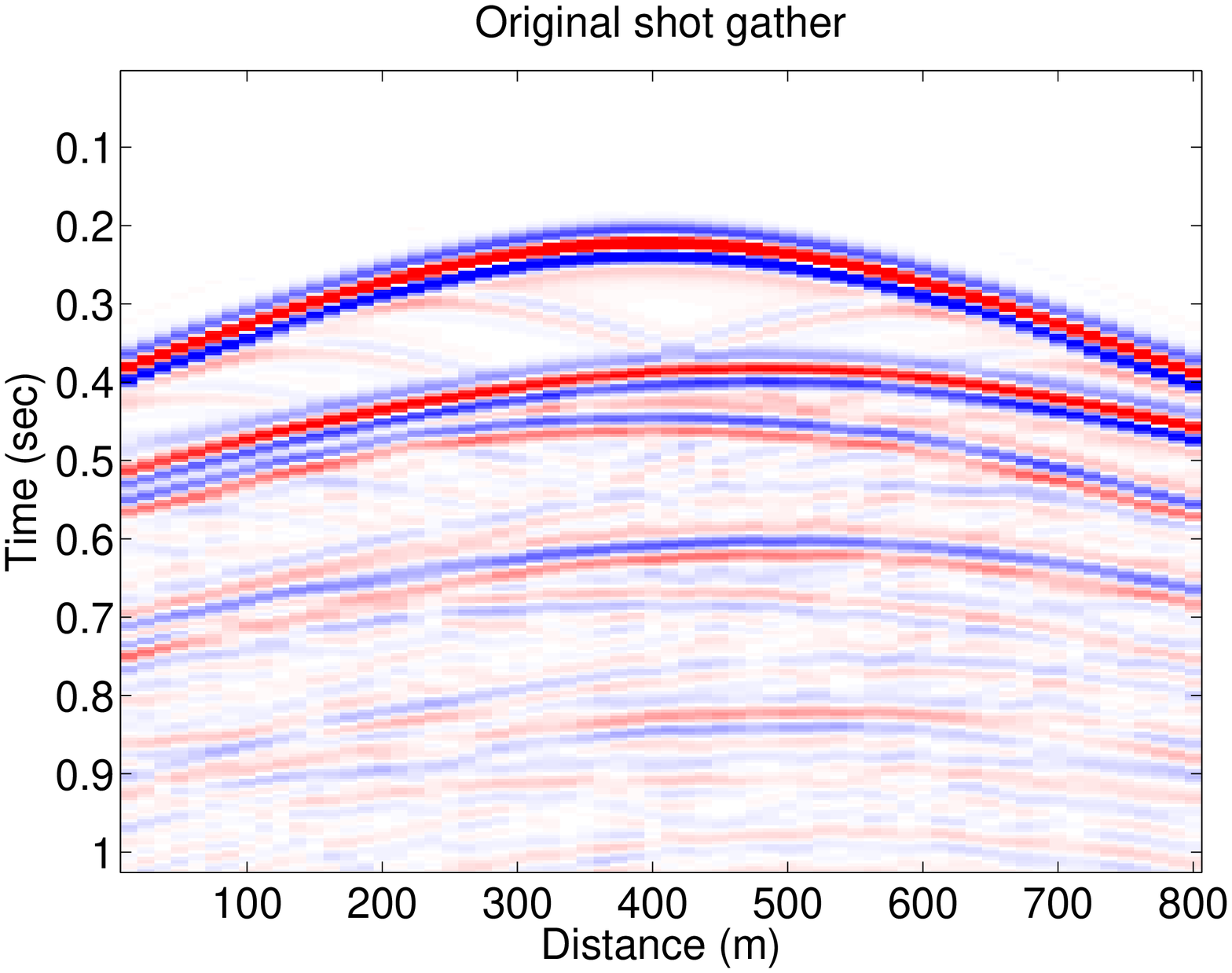}
\label{fig:subfig1}}
\subfloat[][]{
\includegraphics[width=0.48\textwidth]{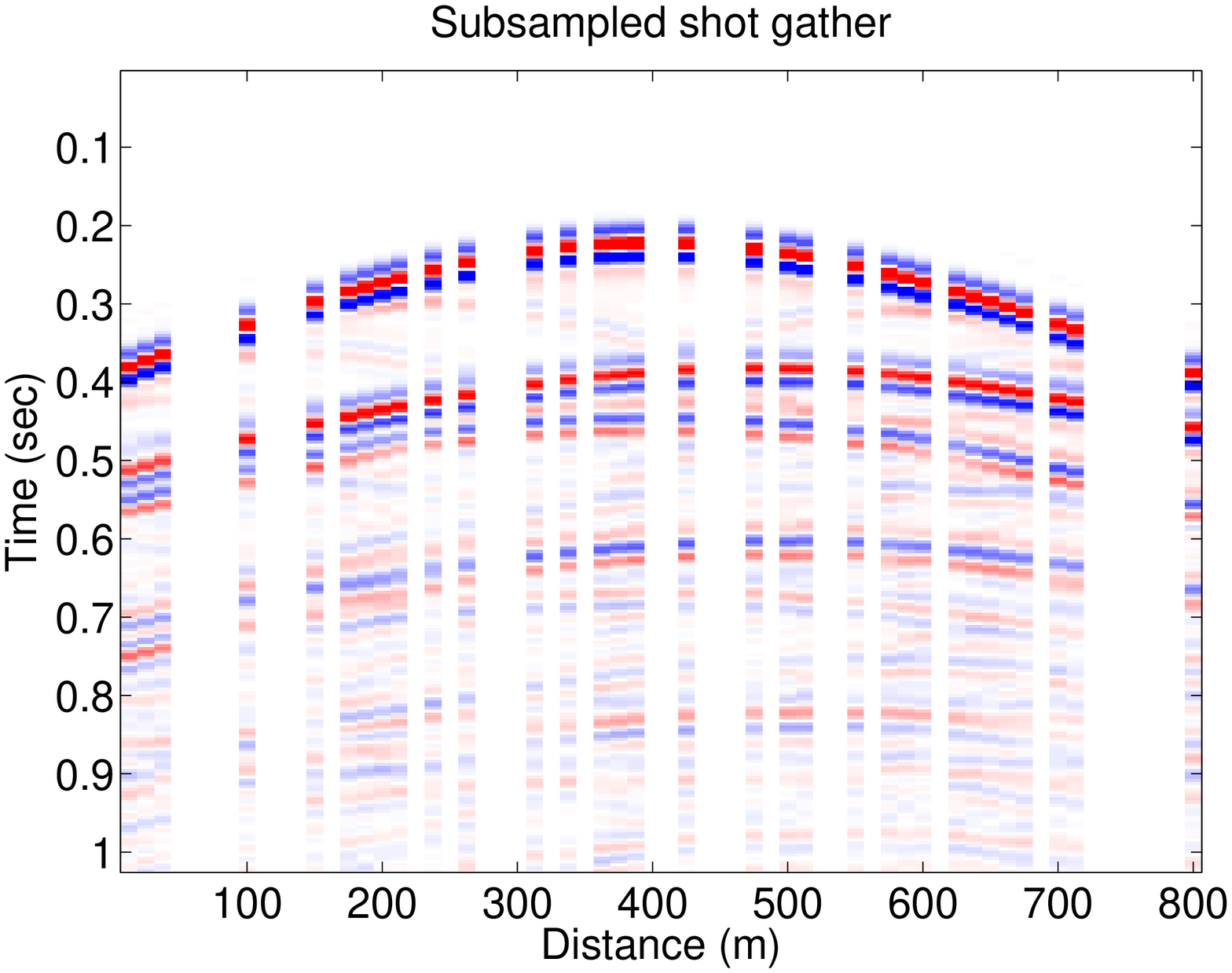}
\label{fig:subfig2}}
\caption{ (a) Shot gather number 32 from the seismic line. (b) Subsampled shot gather using column 32 from the mask in Figure \ref{mask}.b.}
\label{full}
\end{figure*}
Figures \ref{full}.a and \ref{full}.b show a fully sampled and the corresponding subsampled shot gather, respectively. The shot gather corresponds to shot number 32 of the seismic line. Figures \ref{rec1}.a and \ref{rec1}.b show the reconstructed shot gathers using $\ell_1$ minimization and $\ell_p$ minimization, respectively and Figures \ref{rec}.a and \ref{rec}.b show the reconstructed shot gathers using weighted $\ell_1$ minimization and weighted $\ell_p$ minimization, respectively. Furthermore the reconstruction error plots of $\ell_1$ and $\ell_p$ minimization is showed in Figure \ref{recerr1}.a and \ref{recerr1}.b and the reconstruction error plots of weighted $\ell_1$ and weighted $\ell_p$ minimization are shown in Figures \ref{recerr}.a and \ref{recerr}.b.\newline
Figure \ref{SNRshotgather} shows the SNRs of all shot gathers recovered by using regular and weighted and regular $\ell_p$ and $\ell_1$ minimization problems. The plots demonstrate that recovery by weighted $\ell_p$ in the frequency-source-receiver domain is always better than recovery by regular $\ell_p$. In this plot we also see that although recovery by weighted $\ell_p$ minimization is better than regular $\ell_1$ minimization but the results are just a little better than recovery by weighted $\ell_1$ minimization. We believe that similar to the case we see in the noisy compressible case this is an artifact of the algorithm we are using.\newline
\begin{figure*}[!t]
\centering
\subfloat[][]{
\includegraphics[width=0.53\textwidth, height=0.42\textwidth]{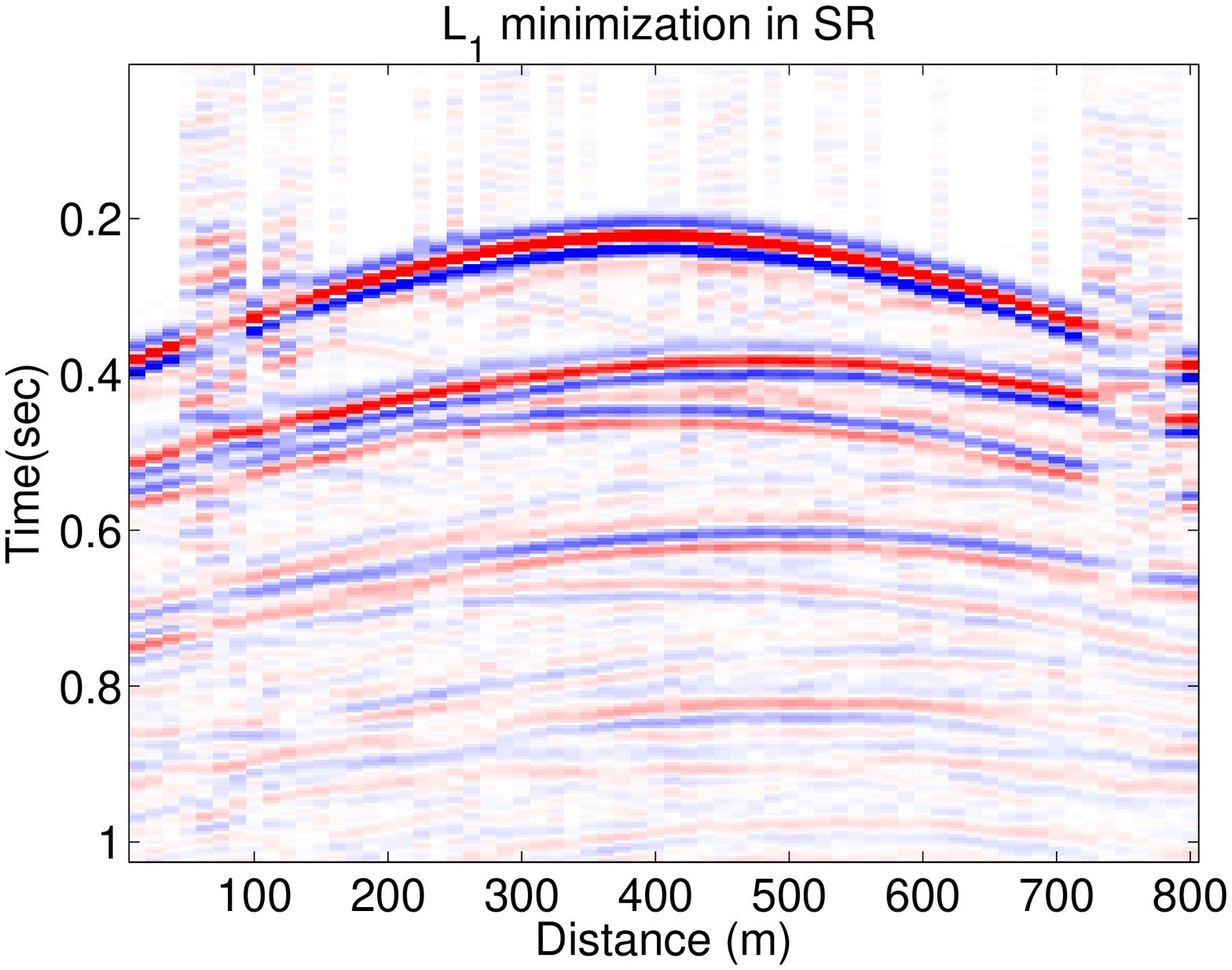}
\label{fig:subfig1}}
\subfloat[][]{
\includegraphics[width=0.53\textwidth, height=0.42\textwidth]{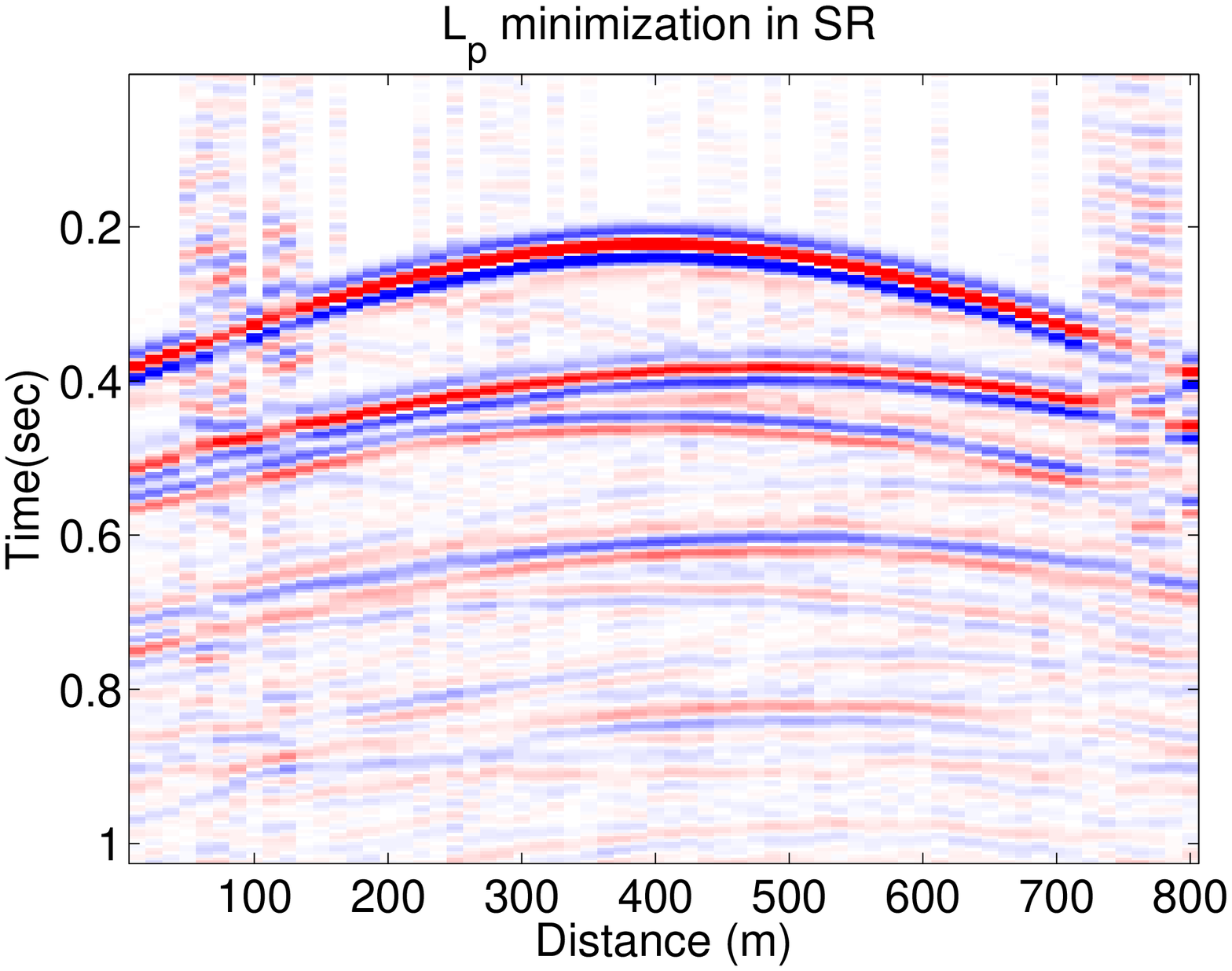}
\label{fig:subfig2}}
\caption{ (a) Recovered shot gather number 32 using $\ell_1$ minimization in the SR domain. (b) Recovered shot gather using $\ell_p$ minimization in the SR domain.}
\label{rec1}
\end{figure*}
\begin{figure*}[!t]
\centering
\subfloat[][]{
\includegraphics[width=0.53\textwidth, height=0.42\textwidth]{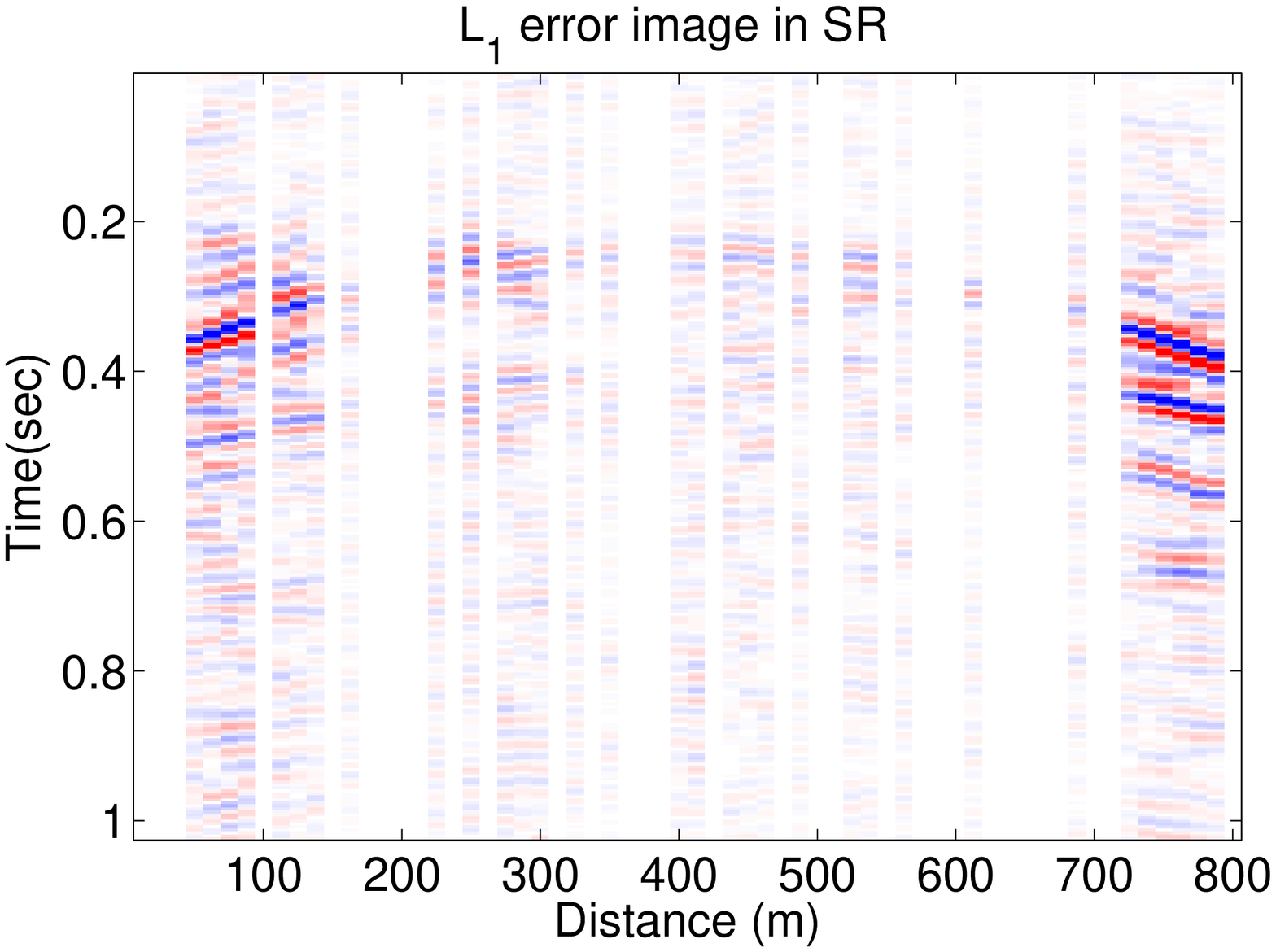}
\label{fig:subfig1}}
\subfloat[][]{
\includegraphics[width=0.53\textwidth, height=0.42\textwidth]{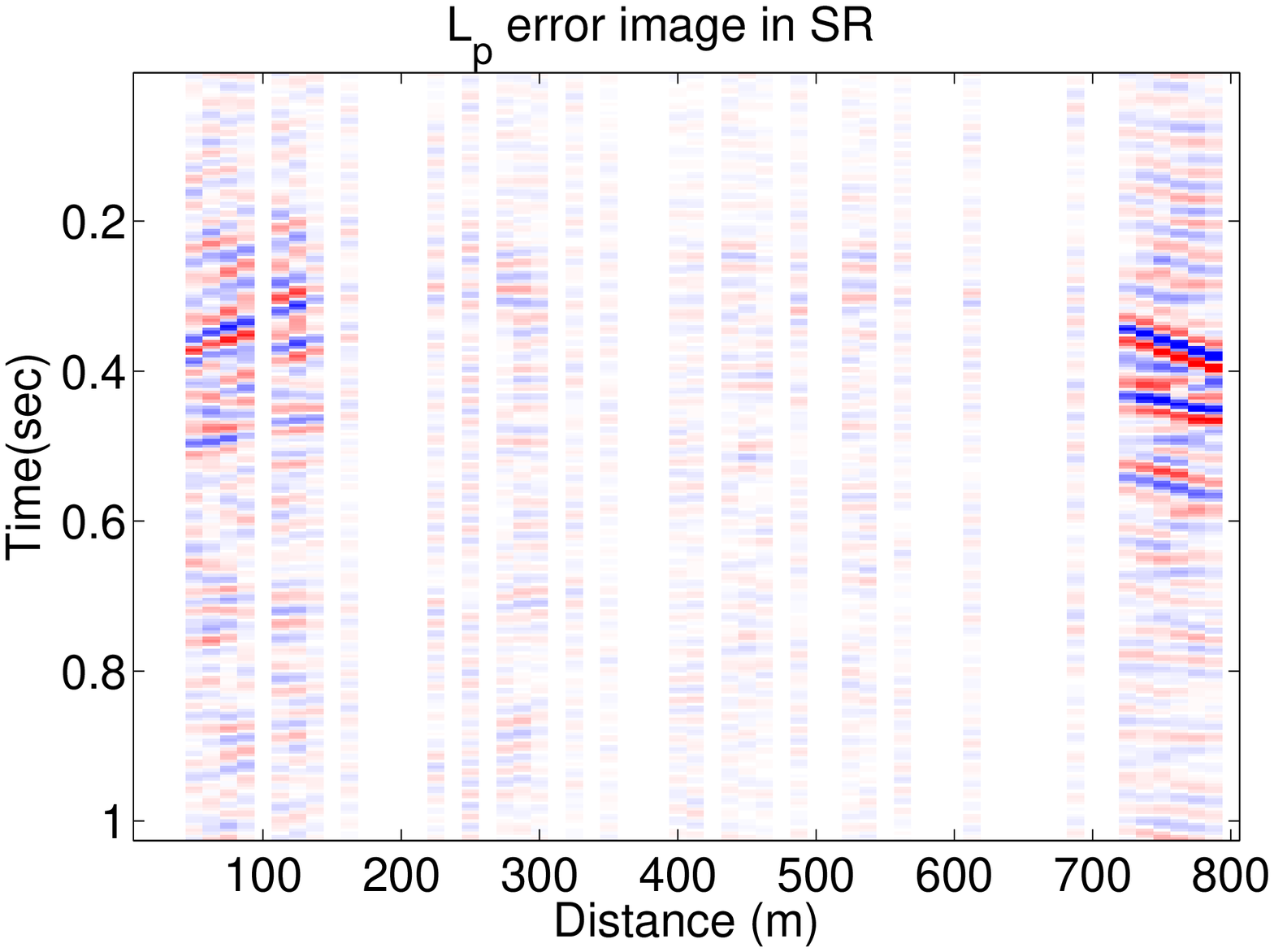}
\label{fig:subfig2}}
\caption{ (a) Error plots showing the difference between the original shot gather and the reconstruction from $\ell_1$ minimization in the source-receiver domain. (b) Error plots showing the difference between the original shot gather and the reconstruction from $\ell_p$ minimization in the SR domain.}
\label{recerr1}
\end{figure*}
\begin{figure*}[!t]
\centering
\subfloat[][]{
\includegraphics[width=0.53\textwidth, height=0.42\textwidth]{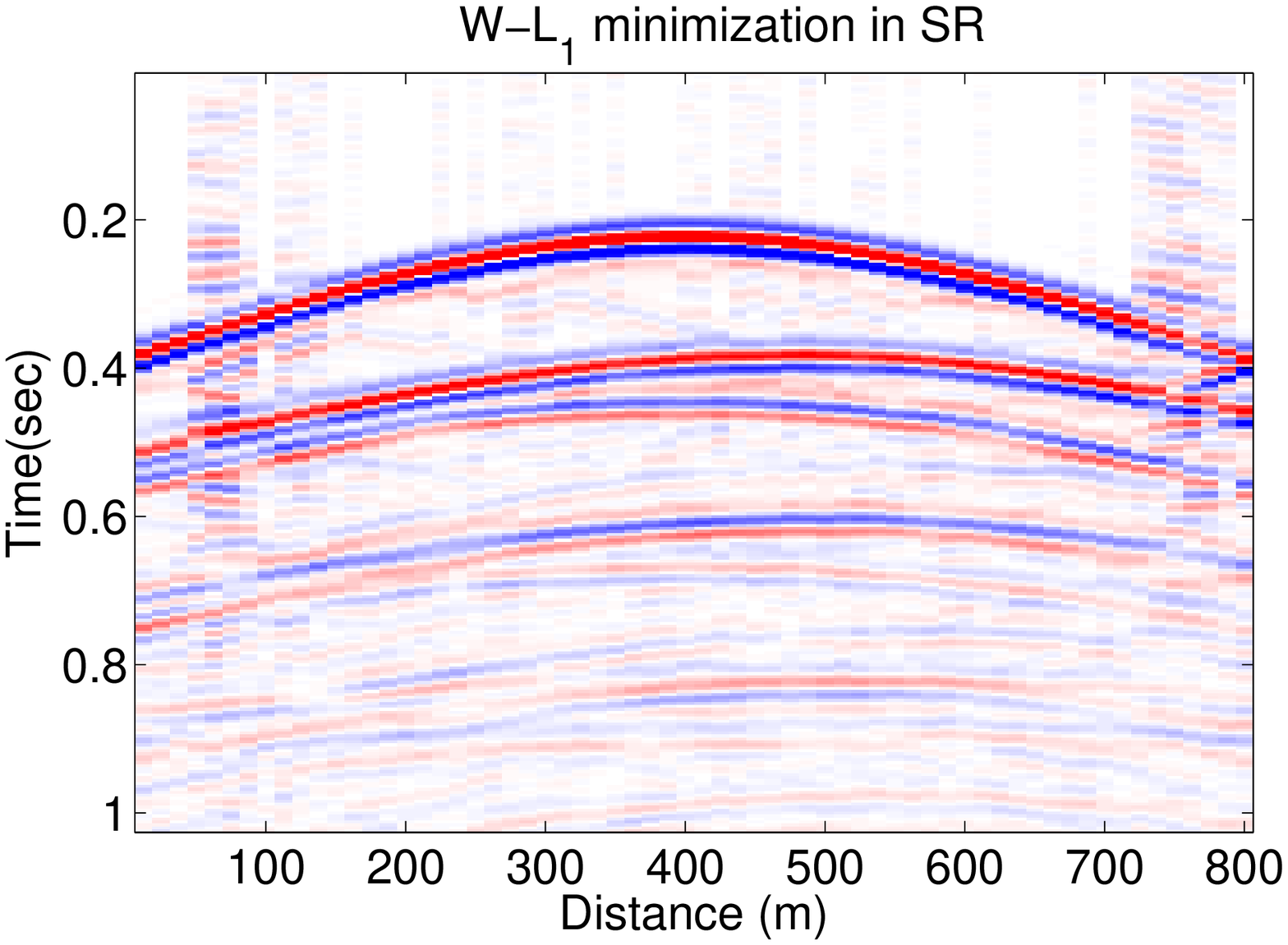}
\label{fig:subfig1}}
\subfloat[][]{
\includegraphics[width=0.53\textwidth, height=0.42\textwidth]{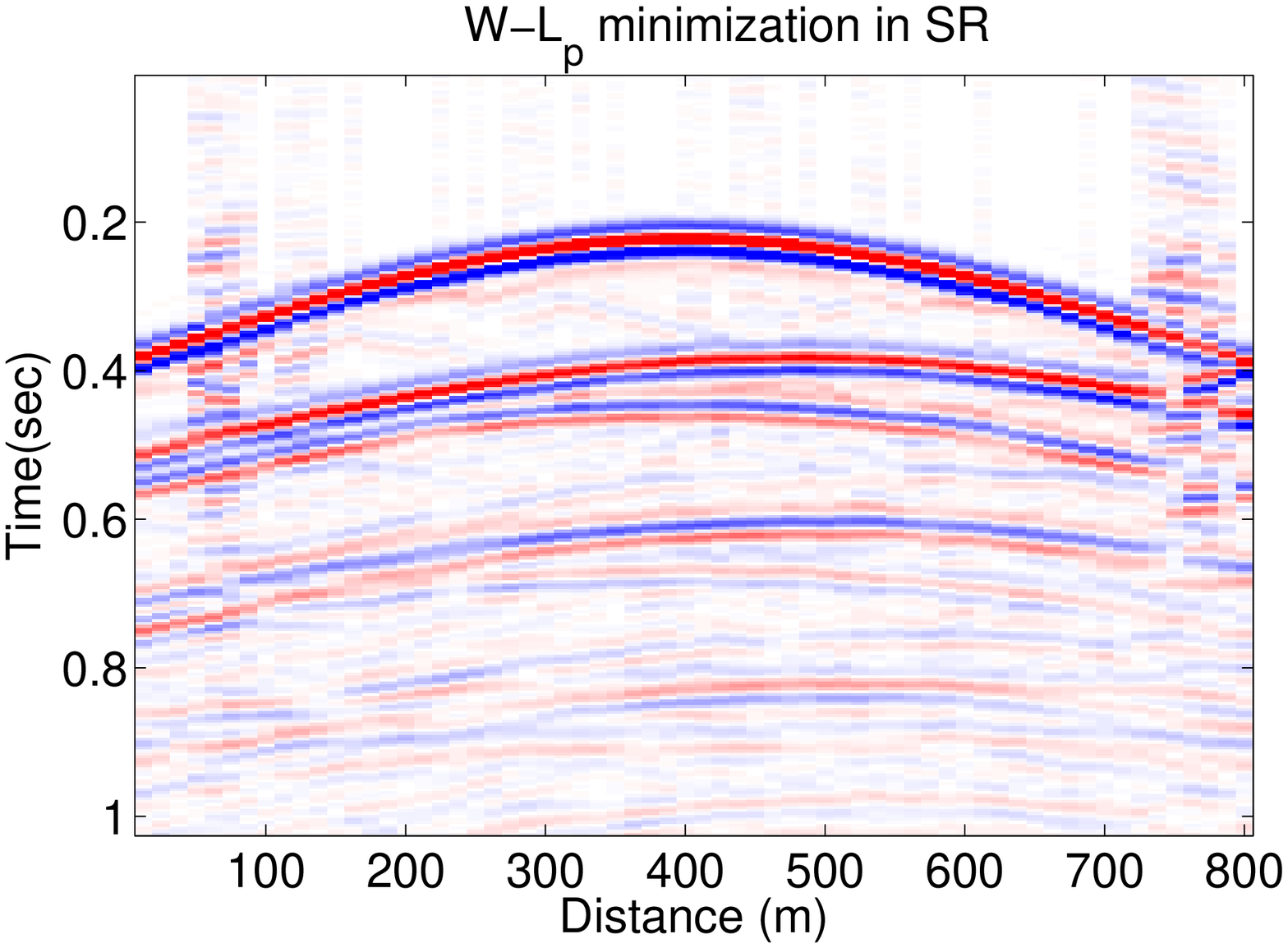}
\label{fig:subfig2}}
\caption{ (a) Recovered shot gather number 32 using weighted $\ell_1$ minimization in the SR domain. (b) Recovered shot gather using weighted $\ell_p$ minimization in the SR domain.}
\label{rec}
\end{figure*}
\begin{figure*}[!t]
\centering
\subfloat[][]{
\includegraphics[width=0.53\textwidth, height=0.42\textwidth]{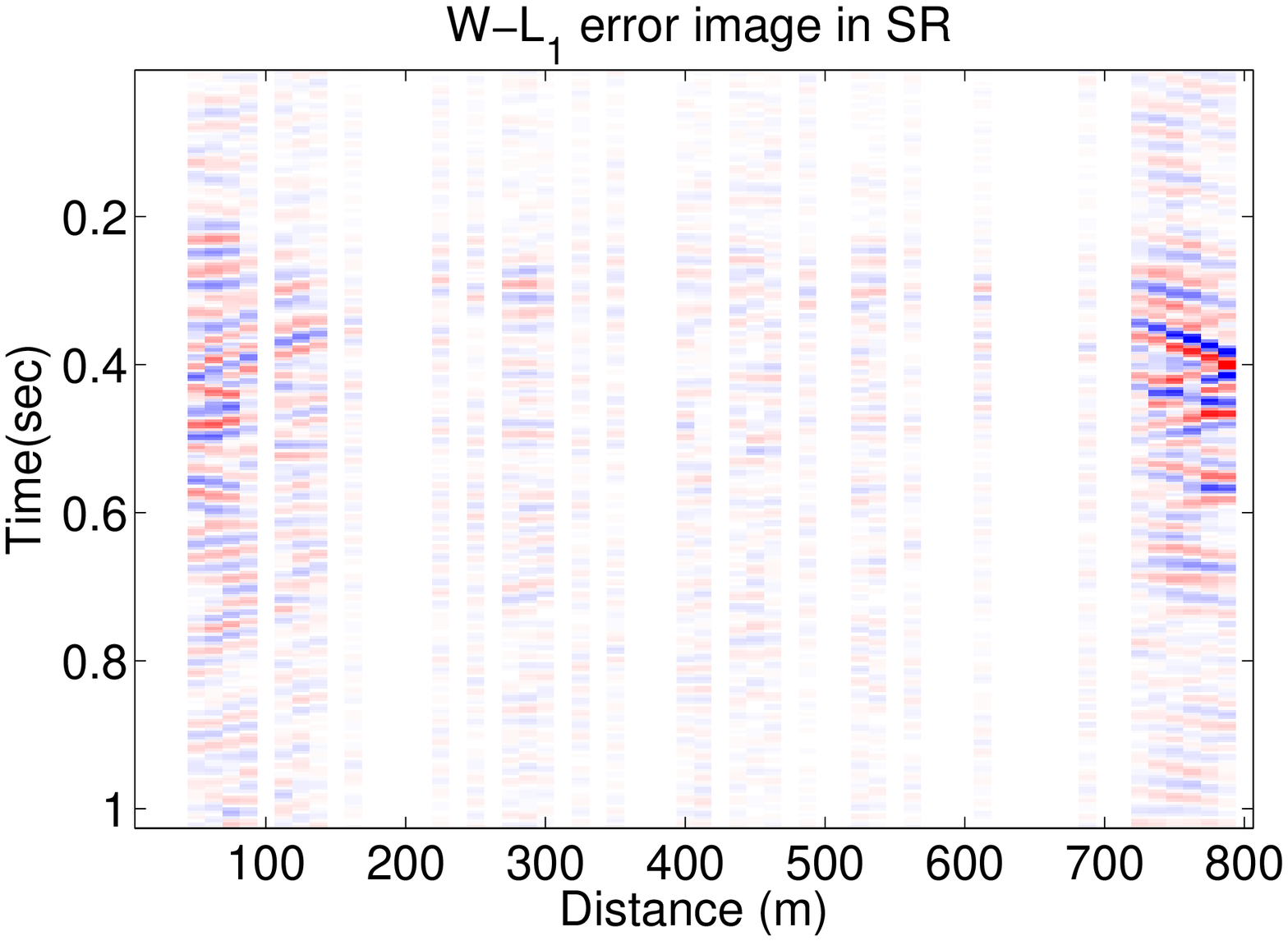}
\label{fig:subfig1}}
\subfloat[][]{
\includegraphics[width=0.53\textwidth, height=0.42\textwidth]{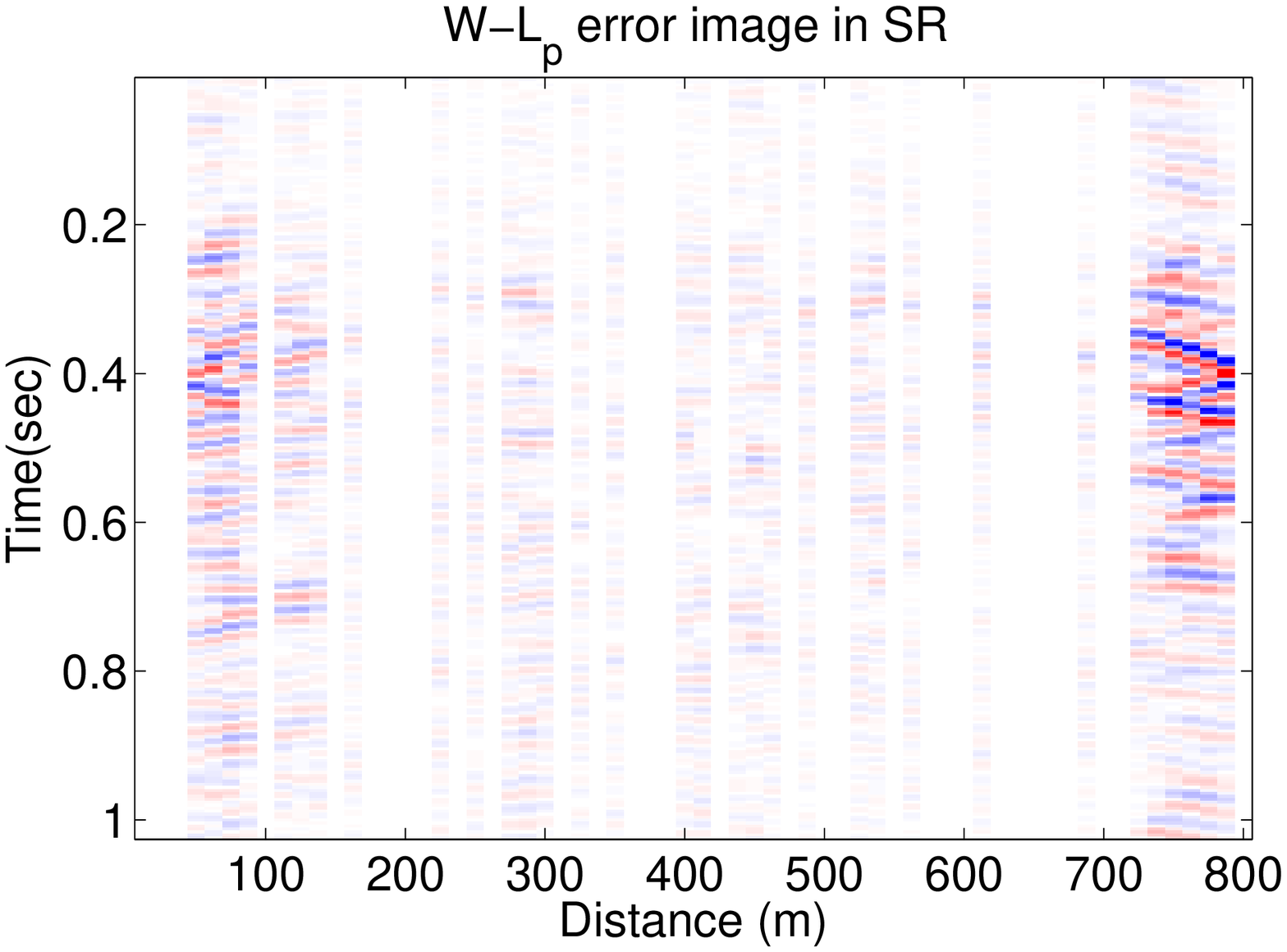}
\label{fig:subfig2}}
\caption{ (a) Error plots showing the difference between the original shot gather and the reconstruction from weighted $\ell_1$ minimization in the source-receiver domain. (b) Error plots showing the difference between the original shot gather and the reconstruction from weighted $\ell_p$ minimization in the SR domain.}
\label{recerr}
\end{figure*}
\begin{figure*}[!t]
\centering
\includegraphics[width=0.8\textwidth]{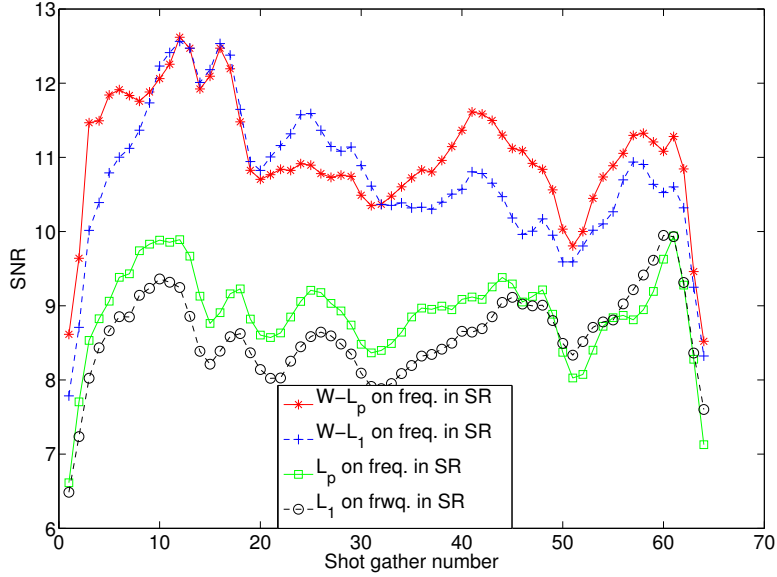}
\caption{Comparison of the SNRs achieved by $\ell_1$, $\ell_p$, weighted $\ell_1$, and weighted $\ell_p$ minimization in recovering shot gathers applied to source-receiver domain}
\label{SNRshotgather}
\end{figure*}
\section{Proof of Theorem \ref{maintheorem1}}
Recall that $\widetilde{T}$, an arbitrary subset of $\{1,2,...,N\}$, is of size $\rho k$ where $0 \leq \rho \leq a$ and $a$ is some number larger than $1$. Let the set $ \widetilde{T}_{\alpha}=T_0 \cap \widetilde{T}$ and $ \widetilde{T}_{\beta}=T_0^c \cap \widetilde{T}$ where, $|\widetilde{T}_{\alpha}|= \alpha \widetilde{T}=\alpha \rho k \ $ and $\alpha +\beta=1$.\newline
Let $x^* = x+h $ be a minimizer of the weighted $\ell_p$ problem. Then 
\begin{figure}[ht]
\centering
\includegraphics[scale=0.8]{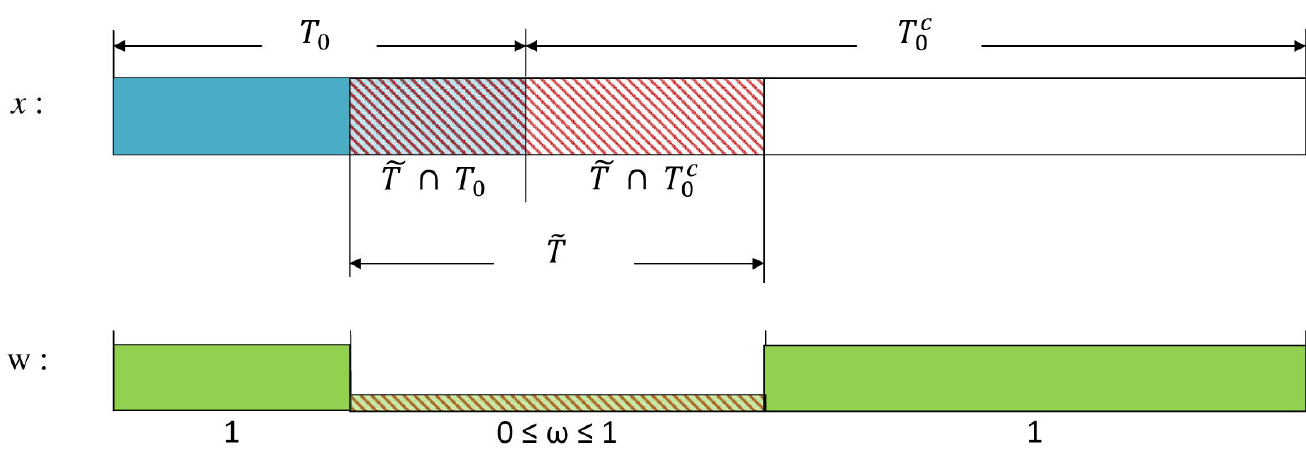}\newline
\caption{Illustration of the signal $x$ and weight vector $\rm{w}$ emphasizing the relationship between the sets $T_0$ and $\widetilde{T}$.}
\end{figure}
$$ \| x+h \|_{p,\rm{w}} \leq \|x \|_{p,\rm{w}} \Rightarrow  \| x+h \|_{p,\rm{w}}^p \leq \|x \|_{p,\rm{w}}^p. $$
Using the weights, we have
$$\omega^p \|x_{\widetilde{T}}+h_{\widetilde{T}} \|_p^p + \|x_{\widetilde{T}^c}+h_{\widetilde{T}^c} \|_p^p \leq \omega^p \|x_{\widetilde{T}} \|_p^p + \|x_{\widetilde{T}^c} \|_p^p. $$\newline
Consequently,
$$\omega^p\|x_{\widetilde{T}\cap T_0}+h_{\widetilde{T}\cap T_0} \|_p^p +\omega^p \|x_{\widetilde{T}\cap T_0^c}+h_{\widetilde{T}\cap T_0^c} \|_p^p +\|x_{\widetilde{T}^c\cap T_0}+h_{\widetilde{T}^c \cap T_0} \|_p^p +\|x_{\widetilde{T}^c\cap T_0^c}+h_{\widetilde{T}^c \cap T_0^c}\|_p^p$$
$$ \leq \omega^p \|x_{\widetilde{T}\cap T_0} \|_p^p +\omega^p \|x_{\widetilde{T}\cap T_0^c} \|_p^p + \|x_{\widetilde{T}^c\cap T_0} \|_p^p +\|x_{\widetilde{T}^c\cap T_0^c}\|_p^p.$$
We use the forward and reverse triangle inequalities to get
$$\omega^p\|h_{\widetilde{T}\cap T_0^c} \|_p^p +\|h_{\widetilde{T}^c\cap T_0^c} \|_p^p \leq \omega^p \|h_{\widetilde{T}\cap T_0} \|_p^p +\|h_{\widetilde{T}^c\cap T_0} \|_p^p +2(\omega^p\| x_{\widetilde{T}\cap T_0^c} \|_p^p +\|x_{\widetilde{T}^c\cap T_0^c}\|_p^p).$$
Adding and subtracting $\omega^p\|h_{\widetilde{T}^c\cap T_0^c}\|_p^p$ to the left hand side and adding and subtracting $\omega^p\|h_{\widetilde{T}^c\cap T_0}\|_p^p+\omega^p\|x_{\widetilde{T}^c\cap T_0^c}\|_p^p$ to the right hand side we get
\normalsize
$$\omega^p\|h_{\widetilde{T}\cap T_0^c}\|_p^p +\omega^p \|h_{\widetilde{T}^c\cap T_0^c}\|_p^p + \|h_{\widetilde{T}^c\cap T_0^c}\|_p^p -\omega^p \|h_{\widetilde{T}^c\cap T_0^c}\|_p^p \leq \omega^p\|h_{\widetilde{T}\cap T_0}\|_p^p +\omega^p \|h_{\widetilde{T}^c\cap T_0}\|_p^p$$
$$+ \|h_{\widetilde{T}^c\cap T_0}\|_p^p -\omega^p \|h_{\widetilde{T}^c\cap T_0}\|_p^p+2(\omega^p\|x_{\widetilde{T}\cap T_0^c}\|_p^p +\omega^p \|x_{\widetilde{T}^c\cap T_0^c}\|_p^p + \|x_{\widetilde{T}^c\cap T_0^c}\|_p^p -\omega^p \|x_{\widetilde{T}^c\cap T_0^c}\|_p^p).$$
\normalsize
Since $\|h_{T_o^c}\|_p^p = \|h_{\widetilde{T}\cap T_0^c}\|_p^p +\|h_{\widetilde{T}^c\cap T_0^c}\|_p^p$ we get
\normalsize
\begin{equation} \label{star}
\begin{aligned}
&\omega^p \|h_{T_0^c}\|_p^p +(1-\omega^p) \|h_{\widetilde{T}^c\cap T_0^c}\|_p^p \leq \omega^p\|h_{T_0}\|_p^p\\
&+ (1-\omega^p) \|h_{\widetilde{T}^c\cap T_0}\|_p^p
 +2(\omega^p\|x_{T_0^c}\|_p^p +(1-\omega^p)\|x_{\widetilde{T}^c\cap T_0^c}\|_p^p).
\end{aligned}
\end{equation}
\normalsize
We also have $\|h_{T_o^c}\|_p^p = \omega^p\|h_{T_o^c}\|_p^p+ (1-\omega^p)\|h_{\widetilde{T}\cap T_0^c}\|_p^p +(1-\omega^p)\|h_{\widetilde{T}^c\cap T_0^c}\|_p^p.$
Combining this with (\ref{star}) we get
\normalsize
\begin{equation} \label{firstone}
\begin{aligned}
&\|h_{T_0^c}\|_p^p \leq \omega^p \|h_{T_0}\|_p^p +(1-\omega^p)(\|h_{\widetilde{T}^c\cap T_0}\|_p^p + \|h_{\widetilde{T}\cap T_0^c}\|_p^p)\\
&+2(\omega^p\|x_{T_0^c}\|_p^p +(1-\omega^p)(\|x_{\widetilde{T}^c\cap T_0^c}\|_p^p).
\end{aligned}
\end{equation}
\normalsize
Define $\widetilde{T}_\alpha:=T_0 \cap \widetilde{T}$. Then $\|h_{\widetilde{T}^c\cap T_0}\|_p^p + \|h_{\widetilde{T}\cap T_0^c}\|_p^p =\|h_{T_0\cup \widetilde{T} \setminus \widetilde{T}_\alpha}\|_p^p$ and from (\ref{firstone})
\begin{equation} \label{one}
\|h_{T_0^c}\|_p^p \leq \omega^p \|h_{T_0}\|_p^p +(1-\omega^p)\|h_{T_0\cup \widetilde{T} \setminus \widetilde{T}_\alpha}\|_p^p+2(\omega^p\|x_{T_0^c}\|_p^p +(1-\omega^p)(\|x_{\widetilde{T}^c\cap T_0^c}\|_p^p).
\end{equation}
Now partition $T_0^c$ into sets of $T_1, T_2 ,...,|T_j|=ak$ for $j\geq 1$, such that $T_1$ is the set of indices of the $ak $ largest (in magnitude) coefficients of $h_{T_0^c}$ and so on. Finally let $T_{01}:=T_0\cup T_1$. Now we can find a lower bound for $\|Ah\|_2^p$ using the RIP condition of the matrix A. We have
\begin{equation}\label{23}
\begin{aligned}
&\|Ah\|_2^p =\|Ah_{T_{01}}+ \sum_{j\geq 2}Ah_{T_j}\|_2^p \geq \|Ah_{T_{01}}\|_2^p -\sum_{j\geq 2}\|Ah_{T_j}\|_2^p\\ &\geq (1-\delta_{ak+|T_0|})^{\frac{p}{2}} \|h_{T_{01}}\|_2^p - (1+\delta_{ak})^{\frac{p}{2}} \sum_{j\geq 2}\|h_{T_j}\|_2^p.
\end{aligned}
\end{equation}
\normalsize
Here we also use the fact that $\|.\|_2^p$ satisfies the triangle inequality for $0<p<1$.\newline
Now we should note that $|h_{T_{j+1}}(l)|^p \leq |h_{T_{j}}(l')|^p$ for all $l\in T_{j+1}$ and $l' \in T_{j}$, and thus $|h_{T_{j+1}}(l)|^p \leq \frac{\|h_{T_j}\|_p^p}{ak}$. It follows that $\|h_{T_j}\|_2^2 \leq (ak)^{1-\frac{2}{p}} \|h_{T_j}\|_p^2$ and consequently
\begin{equation}\label{24}
\sum_{j\geq 2}\|h_{T_j}\|_2^p \leq (ak)^{\frac{p}{2}-1}\sum_{j\geq 1}\|h_{T_j}\|_p^p=(ak)^{\frac{p}{2}-1} \|h_{T_0^c}\|_p^p.
\end{equation}
\normalsize
Using (\ref{24})in (\ref{23}) we get
\begin{equation}\label{endpagetwo}
\|Ah\|_2^p \geq  (1-\delta_{ak+|T_0|})^{\frac{p}{2}} \|h_{T_{01}}\|_2^p - (1+\delta_{ak})^{\frac{p}{2}} (ak)^{\frac{p}{2}-1} \|h_{T_0^c}\|_p^p.
\end{equation}
\normalsize
Next, consider the feasibility of $x^*$ and $x$. Both vectors are feasible, so we have $\|Ah\|_2 \leq 2\varepsilon$. Also note that $|T_0\cup \widetilde{T} \setminus \widetilde{T}_\alpha |=(1+\rho -2\alpha\rho)k$ and $\|h_{T_0}\|_p^p \leq |T_0|^{1-\frac{p}{2}} \|h_{T_0}\|_2^p$. Using these and (\ref{one}) in (\ref{endpagetwo}) we get
\begin{equation}
\begin{aligned}
&(1-\delta_{ak+|T_0|})^{\frac{p}{2}} \|h_{T_{01}}\|_2^p \leq (2\varepsilon)^p + 2(1+\delta_{ak})^{\frac{p}{2}} (ak)^{\frac{p}{2}-1} \left(\omega^p\|x_{T_0^c}\|_p^p +(1-\omega^p)\|x_{\widetilde{T}^c\cap T_0^c}\|_p^p\right) +\\ &(1+\delta_{ak})^{\frac{p}{2}} (ak)^{\frac{p}{2}-1}(\omega^p |T_0|^{1-\frac{p}{2}} \|h_{T_0}\|_2^p+(1-\omega^p)\left((1+\rho -2\alpha\rho)k\right)^{1-\frac{p}{2}}\|h_{T_0\cup \widetilde{T} \setminus \widetilde{T}_\alpha}\|_2^p).
\end{aligned}
\end{equation}
\normalsize
$T_1$ contains the largest $ak$ coefficients of $h_{T_0^c}$ with $a>1$. So $|\widetilde{T} \setminus \widetilde{T}_\alpha |=(1-\alpha )\rho k \leq ak$ then $\|h_{T_0\cup \widetilde{T} \setminus \widetilde{T}_\alpha}\|_2 \leq \|h_{T_{01}}\|_2$.\newline
Defining $E_{\omega}:=(\omega^p\|x_{T_0^c}\|_p^p +(1-\omega^p)\|x_{\widetilde{T}^c\cap T_0^c}\|_p^p)$ and $S_{\omega}:=(\omega^p |T_0|^{1-\frac{p}{2}}+(1-\omega^p)\left((1+\rho -2\alpha\rho)k\right)^{1-\frac{p}{2}})$ and using $\|h_{T_{0}}\|_2 \leq \|h_{T_{01}}\|_2$ we have
\begin{equation}\label{ht01}
\|h_{T_{01}}\|_2^p \leq \frac{(2\varepsilon)^p + 2(1+\delta_{ak})^{\frac{p}{2}} (ak)^{\frac{p}{2}-1} E_{\omega}}
{(1-\delta_{ak+|T_0|})^{\frac{p}{2}}-(1+\delta_{ak})^{\frac{p}{2}} (ak)^{\frac{p}{2}-1} S_{\omega}}.
\end{equation}
\normalsize
To complete the proof denote by $h_{T_0^c}[m]$ the $m$-th largest coefficient of $h_{T_0^c}$ and observe that $|h_{T_0^c}[m]|^p \leq \frac{\|h_{T_0^c}\|_p^p}{m}$. As $h_{T_{01}^c}[m]=h_{T_0^c}[m+ak]$ we have:
\begin{equation}\label{30}
\|h_{T_{01}^c}\|_2^2= \sum_{m \geq ak+1} |h_{T_0^c}[m]|^2 \leq \sum_{m \geq ak+1} (\frac{\|h_{T_0^c}\|_p^p}{m})^{\frac{2}{p}} \leq \frac{\|h_{T_0^c}\|_p^2}{(ak)^{\frac{2}{p}-1}(\frac{2}{p}-1)}.
\end{equation}
\normalsize
The last inequality follows because for $0<p<1$:
$$\sum_{m \geq ak+1} m^{-\frac{2}{p}} \leq \int_{ak}^{\infty} t^{-\frac{2}{p}}dt =\frac{1}{(ak)^{\frac{2}{p}-1}(\frac{2}{p}-1)}.$$
Combining (\ref{30}) with (\ref{one}) we get
\begin{equation}\label{two}
\begin{aligned}
&\|h_{T_{01}^c}\|_2^p \leq  \left((ak)^{\frac{2}{p}-1}(\frac{2}{p}-1)\right)^{-\frac{p}{2}} (\omega^p \|h_{T_0}\|_p^p +\\ &(1-\omega^p)\|h_{T_0\cup \widetilde{T} \setminus \widetilde{T}_\alpha}\|_p^p+2(\omega^p\|x_{T_0^c}\|_p^p +(1-\omega^p)(\|x_{\widetilde{T}^c\cap T_0^c}\|_p^p))).
\end{aligned}
\end{equation}
\normalsize
We showed that $\|h_{T_0\cup \widetilde{T} \setminus \widetilde{T}_\alpha}\|_2 \leq \|h_{T_{01}}\|_2$ and $\|h_{T_{0}}\|_2 \leq \|h_{T_{01}}\|_2.$\newline
Using these in \eqref{two} we get
\begin{equation}\label{ht01c}
\begin{aligned}
&\|h_{T_{01}^c}\|_2^p \leq ((ak)^{\frac{2}{p}-1}(\frac{2}{p}-1))^{-\frac{p}{2}} \ast \left((\omega^p |T_0|^{1-\frac{p}{2}} +(1-\omega^p)((1+\rho -2\alpha\rho)k)^{1-\frac{p}{2}}\right)\|h_{T_{01}}\|_2^p \\ &+ 2\left(\omega^p\|x_{T_0^c}\|_p^p +(1-\omega^p)(\|x_{\widetilde{T}^c\cap T_0^c}\|_p^p))\right).
\end{aligned}
\end{equation}
\normalsize
We can find a bound for $\|h\|_2$ using \eqref{ht01} and \eqref{ht01c}
\begin{equation}\label{h}
\begin{aligned}
\|h\|_2^2=(\|h_{T_{01}}\|_2^p)^{\frac{2}{p}}+(\|h_{T_{01}^c}\|_2^p)^{\frac{2}{p}} \leq \left(\|h_{T_{01}}\|_2^p+\|h_{T_{01}^c}\|_2^p\right)^{\frac{2}{p}}. \\
\end{aligned}
\end{equation}
\small
\begin{equation}\label{C1C2}
\|h\|_2^p \leq \frac{2^p\left(1+\frac{S_{\omega}}{\left((ak)^{\frac{2}{p}-1}(\frac{2}{p}-1)\right)^{\frac{p}{2}}}\right) \varepsilon^p}{(1-\delta_{ak+|T_0|})^{\frac{p}{2}}-(1+\delta_{ak})^{\frac{p}{2}} (ak)^{\frac{p}{2}-1} E_{\omega}}
+\frac{2\left((1+\delta_{a})^{\frac{p}{2}} a^{\frac{p}{2}-1}+\frac{(1-\delta_{(a+1)k})^{\frac{p}{2}}}{\left(a^{\frac{2}{p}-1}(\frac{2}{p}-1)\right)^{\frac{p}{2}}}\right) E_{\omega}}{(1-\delta_{ak+|T_0|})^{\frac{p}{2}}-(1+\delta_{ak})^{\frac{p}{2}} (ak)^{\frac{p}{2}-1} S_{\omega}},
\end{equation}
\normalsize
with the condition that the denominator is positive, equivalently:
\small
\begin{equation}\label{deltap}
\begin{aligned}
\delta_{ak}+\frac{a^{\frac{2}{p}-1}}{(\omega^p+(1-\omega^p)(1+\rho-2\alpha\rho)^{1-\frac{p}{2}})^{\frac{2}{p}}}\delta_{(a+1)k} < \frac{a^{\frac{2}{p}-1}}{(\omega^p+(1-\omega^p)(1+\rho-2\alpha\rho)^{1-\frac{p}{2}})^{\frac{2}{p}}}-1.
\end{aligned}
\end{equation}
\normalsize
\vspace{13pt}
\centerline{ACKNOWLEDGEMENT}
\vspace{13pt}
\noindent This work was supported in part by the Natural Sciences and Engineering Research Council of Canada (NSERC)  Discovery Grant (22R82411), the NSERC Accelerator Award (22R68054) and the NSERC Collaborative Research and Development Grant DNOISE II (22R07504). This research was carried out as part of the SINBAD II project with support from the following organizations: BG Group, BP, BGP, Chevron, ConocoPhillips, Petrobras, PGS, Total SA, WesternGeco, Woodside, Ion, and CGG. 

%
%
%
%
%
%

%
%


%
%
%
%
%
%
%
\bibliography{sparse}
\bibliographystyle{plain}
\end{document}